\documentclass{ijuc}
\usepackage{graphicx}
\usepackage[dvips]{color}       
\usepackage{bm}                 
\usepackage{amsmath}            
\usepackage{amsfonts}
\usepackage{amssymb}
\usepackage{latexsym}           

\begin{document}
%
%

%
\title{Generation of topologically useful entangled states}
%
%
\author{Neil B. Lovett \email{pynbl@leeds.ac.uk} \and Benjamin T. H. Varcoe \email{B.Varcoe@leeds.ac.uk}}


\institute{School of Physics and Astronomy, University of Leeds, Leeds, LS2 9JT,
 United Kingdom}
\date{\today}

%
\maketitle              

\begin{abstract}
Measurement based quantum computation requires the generation of a cluster state (quantum resource) prior to starting a computation. Generation of this entangled state can be difficult with many schemes already proposed. We present an abstract scheme which can create 2D cluster states as a universal resource for quantum computing. We find a linear scaling of grid size with cluster depth. The scheme is also capable of creating more exotic topologies including 3D structures and the unit cell for topological error correction. We note its relevance to the cavity QED scheme in \cite{varcoe06a} although it could be applied to various architectures. 
\end{abstract}


\section{Introduction}
\label{sec:intro}

The promise of quantum computation to provide computation fundamentally faster than current classical computers is dependent on finding both a scaleable architecture, which is experimentally feasible, and also useful algorithms. Much progress has been made over the years since Feynman, \cite{feynman85a}, originally proposed the idea of computing using quantum mechanical laws. Shor's algorithm for factoring numbers with an exponential speed up, \cite{shor97a}, and Grover's for searching an unsorted database with an quadratic speed up, \cite{grover96a}, are probably the two most important. Work on quantum algorithms has progressed rapidly especially with the introduction of algorithms based on quantum random walks \cite{aharonov00a, ambainis01a, farhi98a}, the most notable being the search algorithm of Shenvi, Kempe and Whaley \cite{shenvi02a} and an algorithm for element distinctness from Ambainis \cite{ambainis04c}. In fact, the quantum walk is now a standard technique in quantum algorithm design having recently been proven to be universal for computation \cite{childs09a, lovett10a}. However, designing and building a physically realisable system has been slower due to the degree of control required to maintain coherent systems. Some important progress has been made in varying architectures including ion traps, solid state and optical systems, \cite{wineland02a, taylor05a, knill01a}. One of the hardest challenges experimentally is to create and maintain entanglement at varying points throughout the computation.

Cluster state quantum computation, \cite{raussendorf01a}, is a different paradigm in quantum computation than the circuit model. It is also known as measurement based quantum computation (MBQC). In MBQC, a general graph state is produced and then each site is entangled with its neighbour(s) by a controlled phase operation (C-Phase), eq.~(\ref{cphase}), entangling the two together as
 \begin{equation}
C_{phase} = \left (
\begin{matrix}
1 & 0 & 0 & 0 \\ 0 & 1 & 0 & 0 \\ 0 & 0 & 1 & 0 \\ 0 & 0 & 0 & -1 \end{matrix} \right ).
\label{cphase}
\end{equation}
In the case of two qubits they are firstly prepared in the $| + \rangle$ state,

\begin{align}
|\psi \rangle &= \frac{1}{\sqrt{2}}(| 0_{1} \rangle + | 1_{1} \rangle) \otimes  \frac{1}{\sqrt{2}}(| 0_{2} \rangle + | 1_{2} \rangle),\nonumber \\
 | \psi \rangle &= \frac{1}{2} (| 00 \rangle + | 01 \rangle + | 10 \rangle + | 11 \rangle).
\label{initialstate}
\end{align}
The C-Phase entangling operation is then applied between the two qubits to leave the resultant entangled state,

\begin{equation}
C_{phase} | \psi \rangle = \frac{1}{2}(| 00 \rangle + | 01 \rangle + | 10 \rangle - | 11 \rangle).
\label{endstate}
\end{equation}
Single qubit rotations and measurements are used to progress the computation. These measurements are based on the previous results which are fed forward. As all the entanglement is generated when the graph state is produced, no entanglement needs to be generated during the computation, which experimentally is challenging. A cluster state is just a specific graph state, a square lattice. When first introduced it was hoped that this lack of ad-hoc entanglement would mean the experimental implementation would be much easier. However, this has not been the case, although work by Rudolf et al. has shown it to be feasible, \cite{zeilinger05a}. Since its conception, many other schemes for cluster state preparation have also been introduced, \cite{cho05a,dong06a,gonta09a,wunderlich09a}. 

Recently, there has been a stimulus in work on cluster state generation using photons, \cite{devitt07a,hollenberg08a, devitt09a,ionicioiu09a}, and also topological error correction in cluster states, \cite{raussendorf06a,raussendorf07a,raussendorf07b,bravyi07a,fowler08a}. The photonic module, \cite{devitt07a}, is essentially a `plug and play' cluster state generator. It can deterministically create cluster states and also uses the mobility of the photons to enable any output to be connected to any input. In addition to generating cluster states, another focus of the work is on topological error correction. Raussendorf and Goyal introduced the concept of using multiple qubits to encode one logical qubit in an attempt to make it fault tolerant, \cite{raussendorf07b}. This is in contrast to the traditional cluster state in which the qubits are not protected from loss channels or errors in the system.
 
Due to the recent work on topological error correction and the idea of MBQC, the motivation for this work was to develop a scheme to produce a universal resource for MBQC. We also wanted a way to prepare these states which could be scaled up easily. We numerically modelled states we could theoretically prepare. We found we could create many interesting topologies with various possible applications. The recent work by Elham Keshafi et al. \cite{kashefi09a} on ancilla driven quantum computing has unusual `twisted graph states' as a resource which our scheme could be used to create. The unit cell for topological error correction, \cite{raussendorf07a}, can also be created (with some Z measurements to remove qubits). As a cluster state (or graph state) is just a mathematical object, we review some basic graph theory used here in Sec.~\ref{sec:graph}. We then introduce our basic scheme and the states we are able to prepare in Sec.~\ref{sec:scheme}. We develop it further in Sec.~\ref{sec:exscheme} before discussing its benefits and applications.

\section{Graph Theory}
\label{sec:graph}

\begin{figure}[bt]
\centering
\includegraphics[scale=0.35]{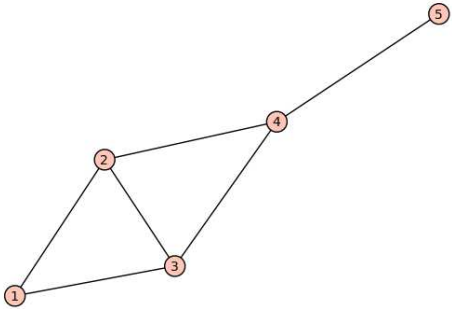}
\caption{An example of a simple general graph with five vertices and six edges. It is undirected, unweighted, connected and has vertices of varying degree.}
\label{graph}
\end{figure}

A general graph is an ordered pair, $G = (V,E)$, where $V$ is a set of vertices and $E$ is a set of edges. The edges in the set $E$ are all described as unordered pairs relating two vertices, $(e_{1},e_{2}) \in V$. The number of vertices in a graph, $|V|$, is the order of the graph and the number of edges, $|E|$, is it's size. In this, the most general description for a graph, the edges are unordered and as such there is no orientation to the graph. This is known as an undirected graph, whereas one with an ordered pair of edges is a directed graph, where a specific direction is given by the edge. The graphs we produce in our scheme are all undirected such as the one shown in fig.~\ref{graph}.
In the scheme we propose (detailed in Sec.~\ref{sec:scheme}), we assume the C-Phase operation, eq.~(\ref{cphase}), can be implemented perfectly and that each pair of qubits is maximally entangled. As such, the edges in our graphs will therefore all be the same, or of the same weight. In this case, our graphs are unweighted. However, if there was some error introduced and some pairs were less entangled than others, this could be modelled by turning our graphs into weighted ones. Weighted graphs have some weighting attached to each edge, which can then be interpreted in different ways. In some optimization problems, this could be a distance or time to travel from vertex to vertex. All the graphs we produce are known as simple graphs. A simple graphs is an undirected, unweighted graph with no self loops (an edge starting and ending at the same vertex) and a maximum of one edge between any two distinct vertices. 

One important feature of graph theory we will use in this work is the notion of connectivity. A graph is said to be connected if there is a sequence of edges from any vertex to any other vertex. In a diagram, a disconnected graph would look like two separate entities. However, the set of vertices and edges of all parts are considered one graph. In our scheme, this is important as it determines whether we have a single copy of the graph produced or multiple ones. The degree of a vertex refers to the measure of adjacent connectivity. The number of edges incident on a vertex is the degree or valency of a vertex. This degree relates to the number of connections made between vertices in the graphs produced in our scheme. Cluster states are just a specific type of simple graph in which only nearest neighbours are connected. This creates a 2D lattice where each internal vertex is of degree four.
\section{Scheme}
\label{sec:scheme}
 
We now present our scheme for the generation of graph states which would be useful in quantum information processing. The scheme we describe is abstract and we do not initially define an architecture in which this could be physically realised. We show some of the useful states we can produce (by numerical simulation) and then discuss the drawbacks of this scheme.

\subsection{Scheme}

\begin{figure}[!t]
\centering
\includegraphics[scale=0.7]{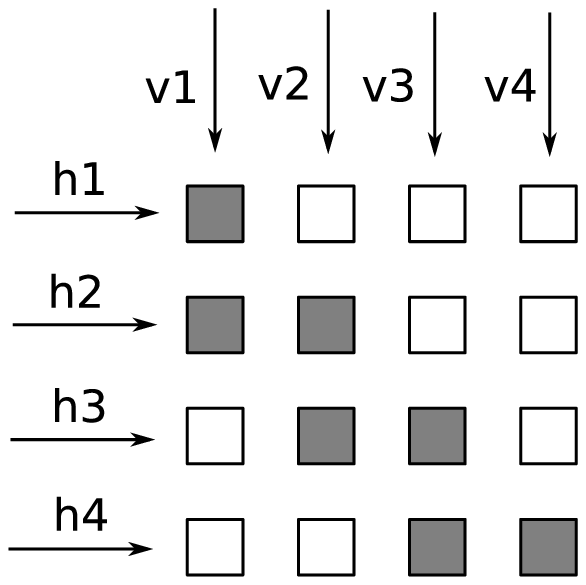}
\caption{Structure we use to create our graph states. We imagine vertices moving in the direction of the arrows advancing by one site in the grid for each timestep. A dark site indicates it is active and so an edge is formed when two vertices pass in the same timestep.}
\label{grid}
\end{figure}

Consider a small grid of sites as in Figure \ref{grid}. We imagine vertices of a graph moving across this grid horizontally and vertically by one site in the grid for each arbitrary timestep. If two of these vertices meet at a site we say a link or edge is formed between them. This edge represents the C-Phase entanglement generated between the two vertices which represent qubits. The vertices enter the grid in the direction of the arrows as a stream of atoms, one entering for each arbitrary timestep. We extend this further and say that each of these sites could be active or inactive, forming an edge only when the site is active. An active site is considered to perform the C-Phase operation between the two vertices (qubits). Therefore instead of creating edges between static vertices, we build our graph states by moving the vertices through a set of `edge joining' interaction regions, which can be switched on or off. This could be a grid of collisional cavities through which atomic beams are passed and entangled together as proposed by Blythe and Varcoe, \cite{varcoe06a}. 

\begin{figure}[!t]
\centering
\includegraphics[scale=0.65]{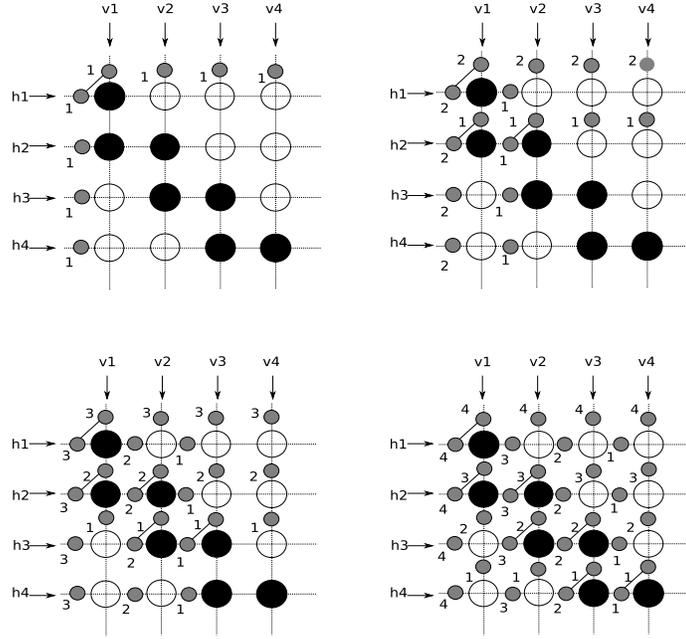}
\caption{First four timesteps of a 4x4 grid. The vertices enter in the direction of the arrows. The black circles indicate active collision sites wheras white indicates the site is switched off.}
\label{fillgrida}
\end{figure}

As we have introduced the concept of the vertices moving with each timestep, we must also address the notion of when (in time) a vertex enters the grid. We do this by assigning a generation to each timestep. Using the grid in fig.~\ref{grid} the vertices entering the first row of the grid will be labelled as $h_{1}g_{4}-h_{1}g_{3}-h_{1}g_{2}-h_{1}g_{1}$. This implies that generation 1 passes into the grid first and so therefore after four timesteps would be at the site in the top right hand corner of the grid. Using this labelling, it is easy to keep track of which generations of vertices are interacting together. We show how the vertices interact with the active sites and different generations in fig.~\ref{fillgrida}. The mix of generations forming edges is obviously highly dependent on the size of the grid and also which sites are active. This dependency and our ability to change these factors allows various different structures to be created. Some of the structures produced by numerical simulation are shown next along with the grid pattern of active sites required to produce them.

\begin{figure}[!t]
\centering
\includegraphics[width=0.45\columnwidth]{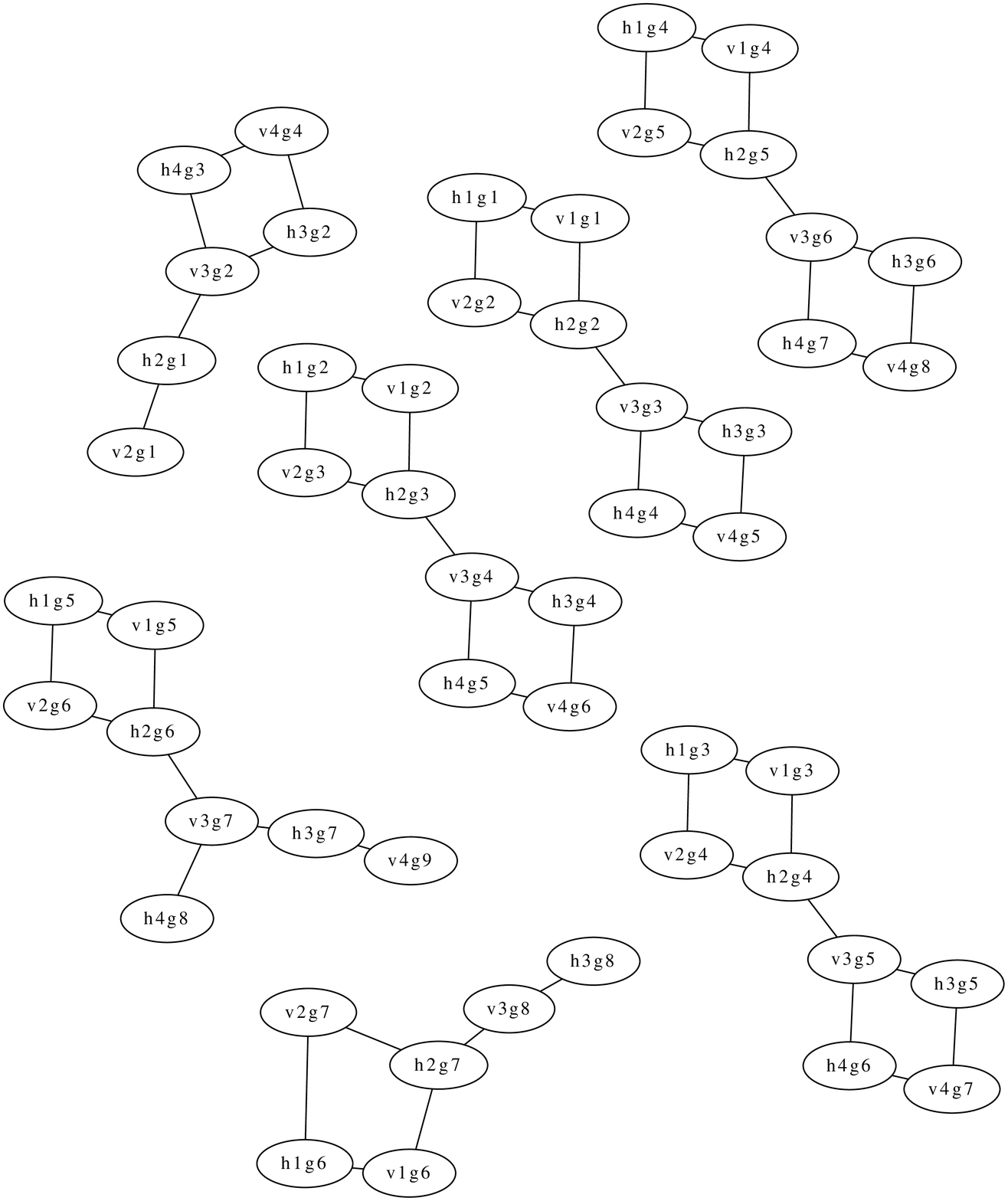}
\line(0,1){180}
\includegraphics[width=0.45\columnwidth]{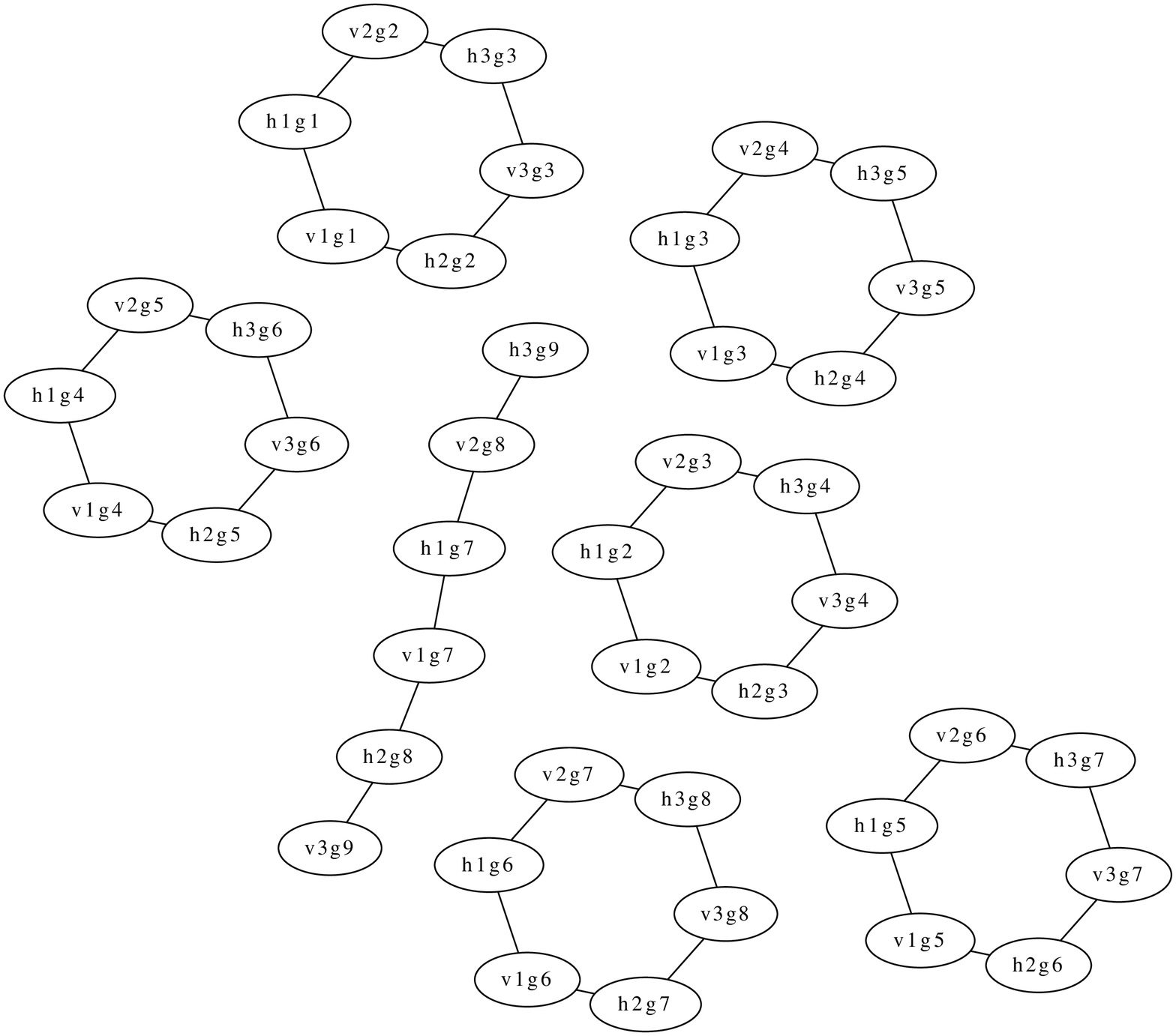}
\caption{Multiple structures produced after 10 timesteps. The incomplete structures can easily be seen. Left - Full set of structures produced from the grid in fig.~\ref{grid2}. Right - Full set of structures produced from the grid in fig.~\ref{grid6}.}
\label{multiple}
\end{figure}

Due to the vertices moving across the grid as a stream, we find that we get a number of the same structures created. The actual number produced depends on which sites are active and the number of timesteps taken. This scales as $t-(\sqrt{N}-1)$ where $t$ is the number of timesteps and $N$ is the number of sites in the grid. However, some of these structures are incomplete. These come about from the vertices that have only partially traversed the grid when the number of timesteps are completed. These multiple and incomplete structures are shown in fig.~\ref{multiple}. In the work  by Blythe and Varcoe, \cite{varcoe06a}, the sequence of atoms is pulsed at specific times to allow the creation of one structure as opposed to a continuous stream here.

\subsection{States produced}

\begin{figure}[!htb]
\centering
\includegraphics[scale=0.55]{grid1a.eps}
\includegraphics[scale=0.18]{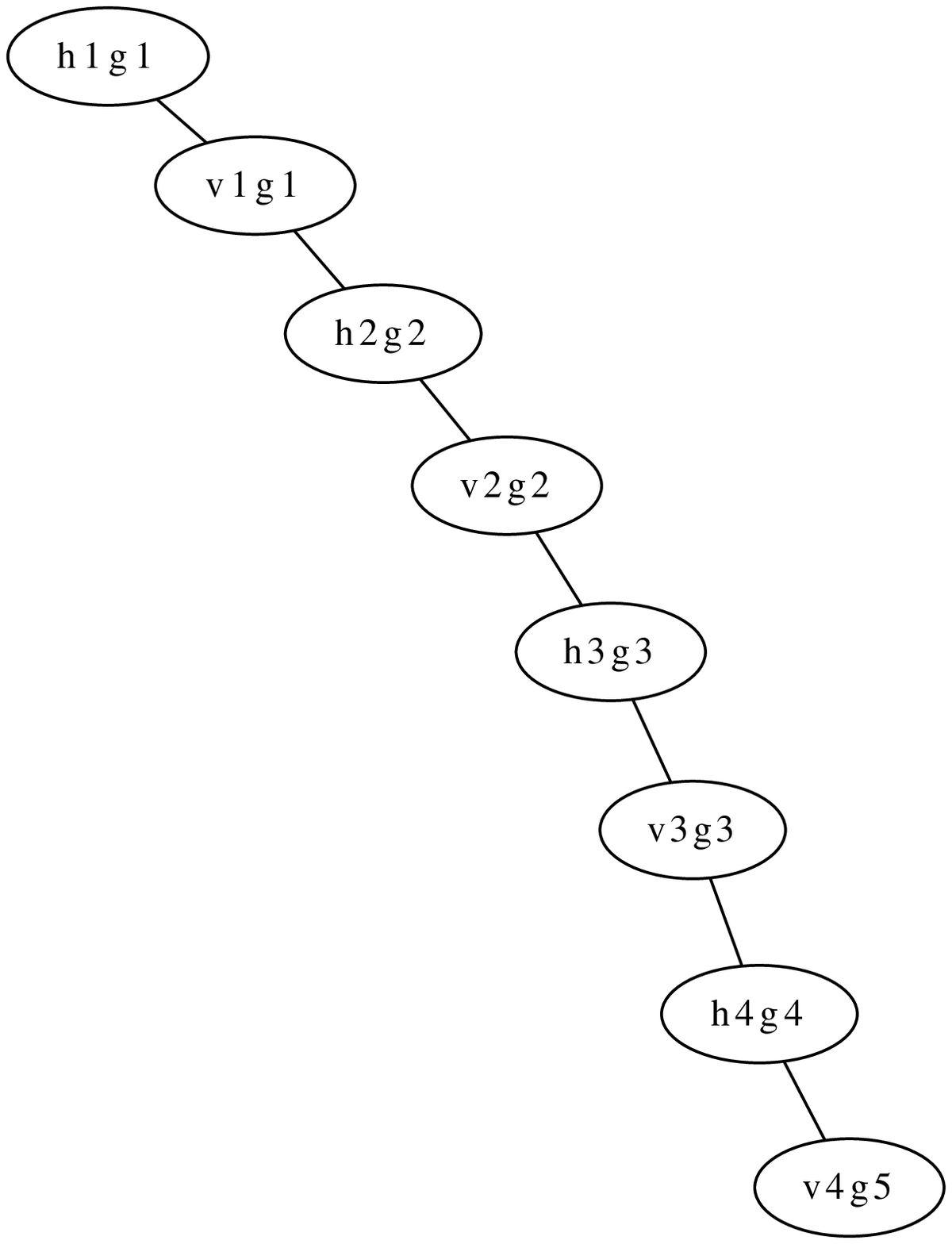}
\caption{Grid of active sites and structure produced. The structure produced is a 1D lattice.}
\label{grid1}
\end{figure}

\begin{figure}[!htb]
\centering
\includegraphics[scale=0.55]{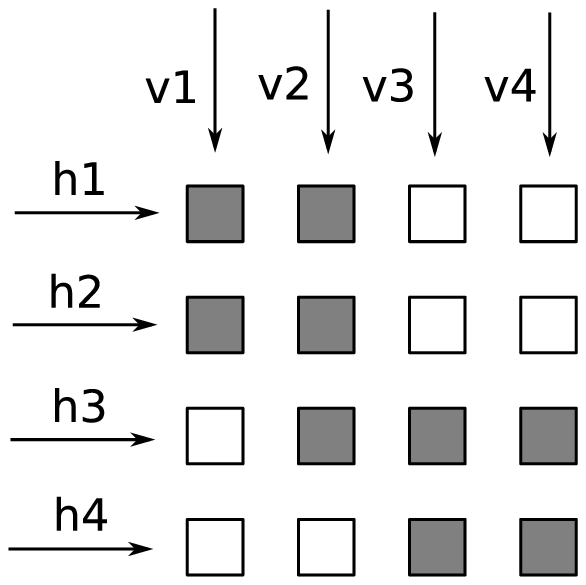}
\includegraphics[scale=0.25]{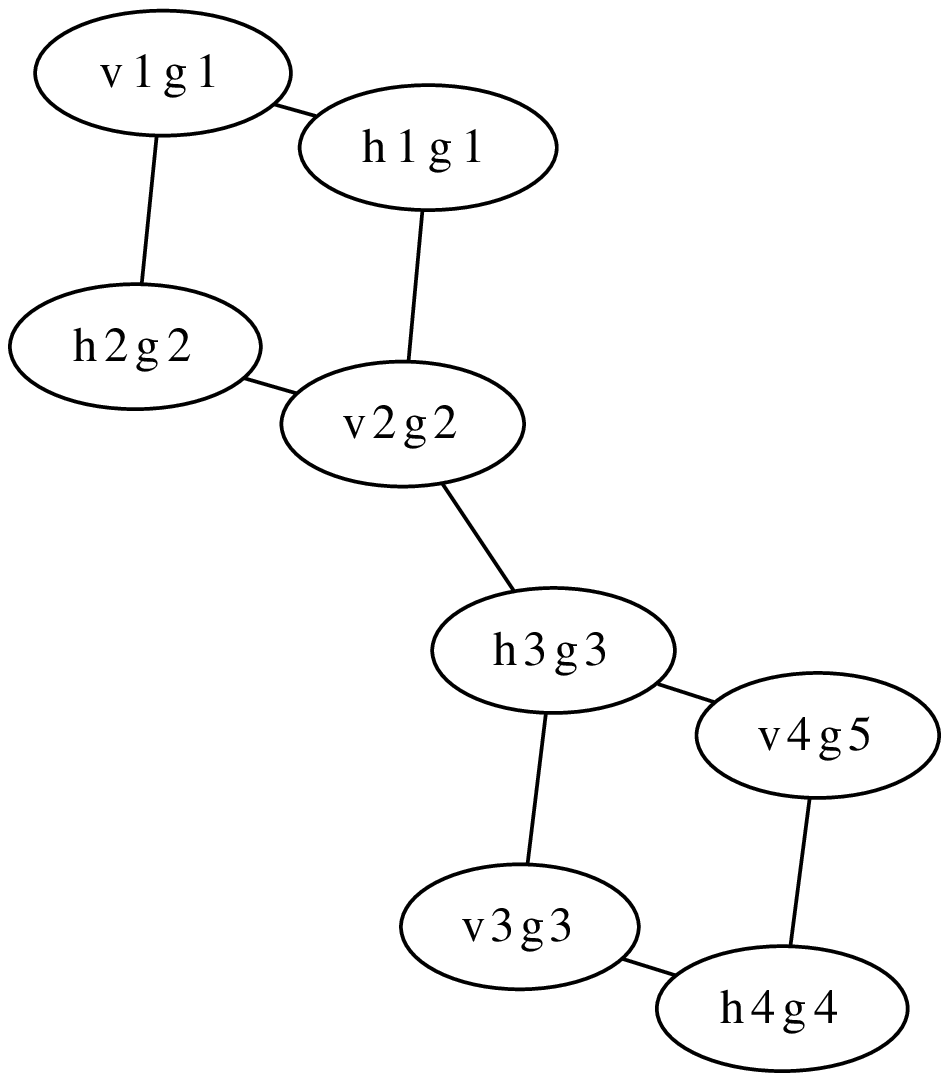}
\caption{Grid of active sites and structure produced. The structure produced is the formation of a 2D lattice but not all the bonds have been formed.}
\label{grid2}
\end{figure}

\begin{figure}[!htb]
\centering
\includegraphics[scale=0.55]{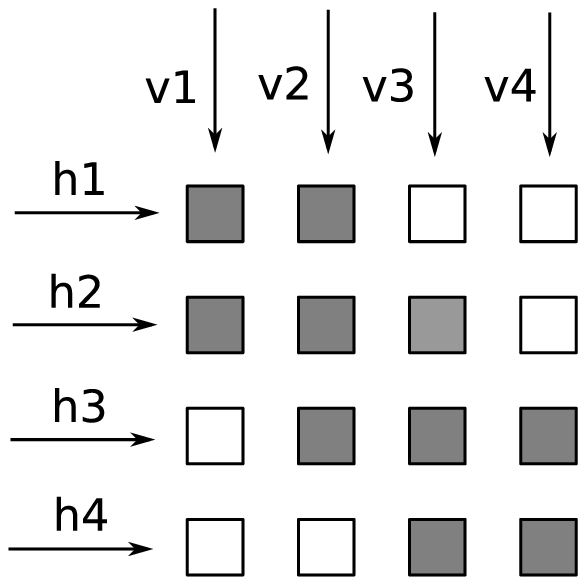}
\includegraphics[scale=0.25]{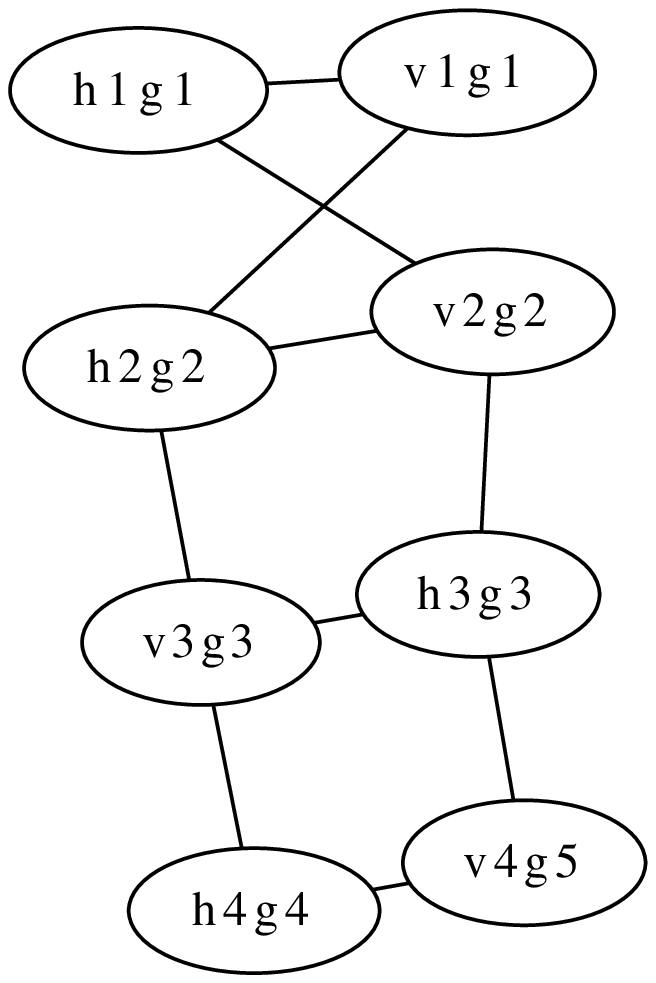}
\caption{Grid of active sites and structure produced. The structure produced is a 2D lattice of depth two.}
\label{grid3}
\end{figure}

\begin{figure}[!htb]
\centering
\includegraphics[scale=0.6]{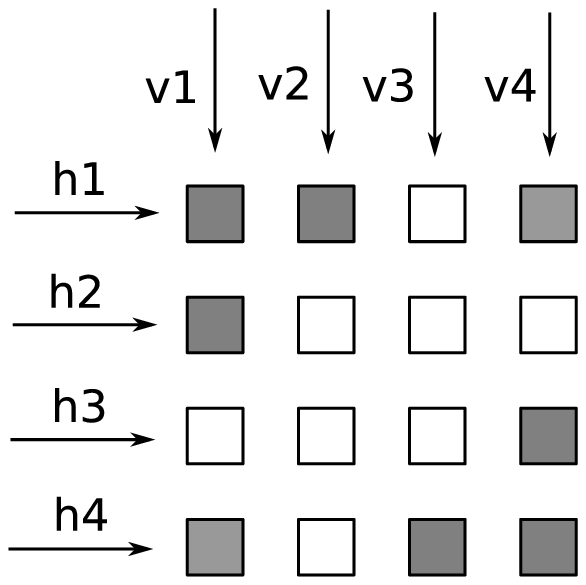}
\includegraphics[scale=0.3]{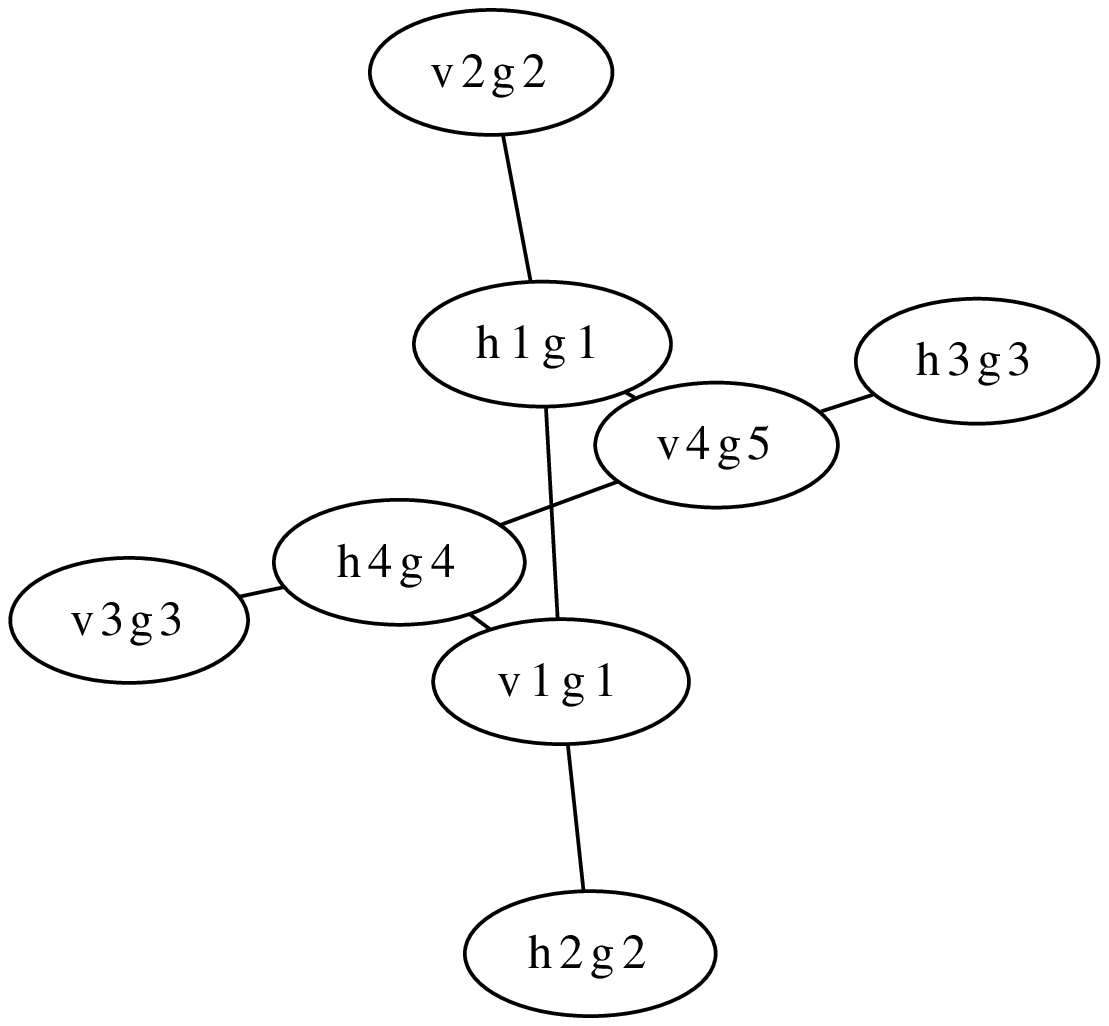}
\caption{Grid of active sites and structure produced. The structure produced is a square lattice with an additional link on each vertex.}
\label{grid5}
\end{figure}

\begin{figure}[!htb]
\centering
\includegraphics[scale=0.55]{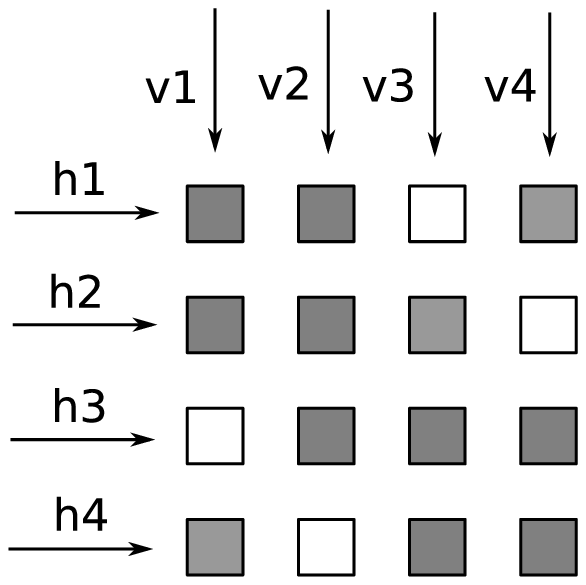}
\includegraphics[scale=0.25]{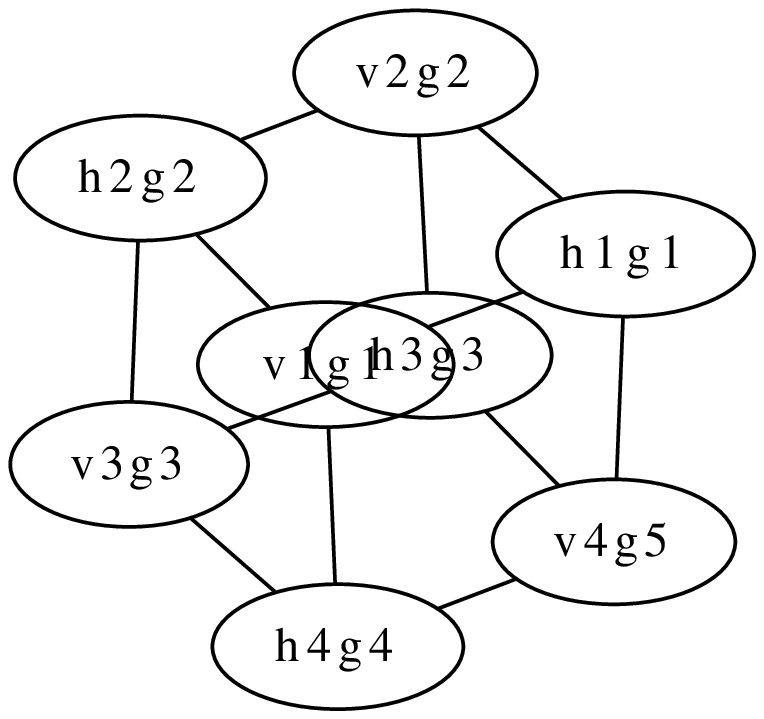}
\caption{Grid of active sites and structure produced. The structure produced is a cube.}
\label{grid4}
\end{figure}

\begin{figure}[!htb]
\centering
\includegraphics[scale=0.55]{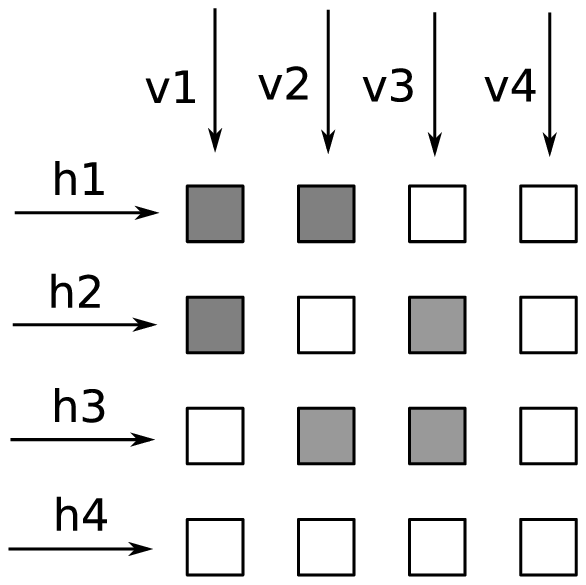}
\includegraphics[scale=0.25]{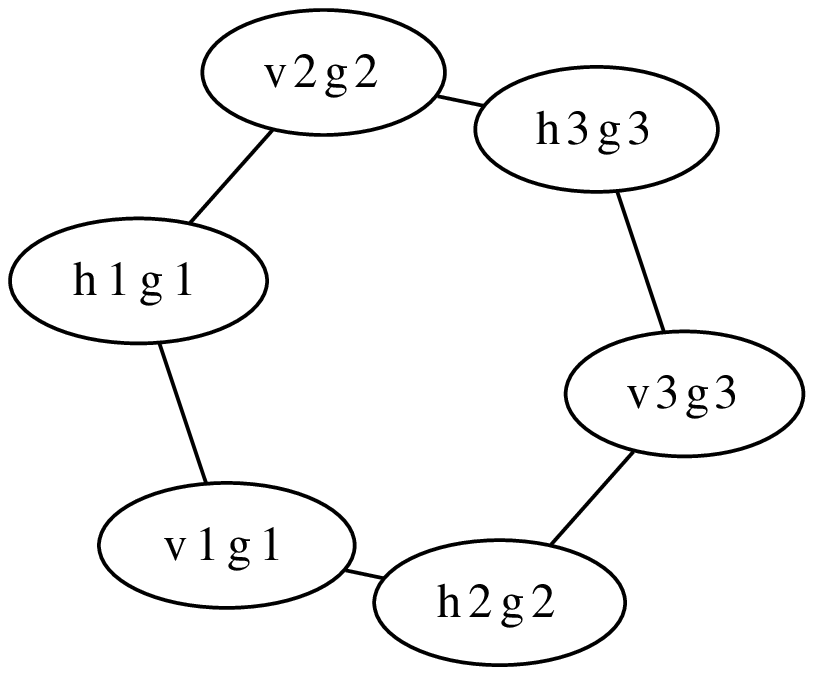}
\caption{Grid of active sites and structure produced. The structure produced is a single cell of a hexagonal lattice.}
\label{grid6}
\end{figure}

\begin{figure}[!h]
\centering
\includegraphics[scale=0.35]{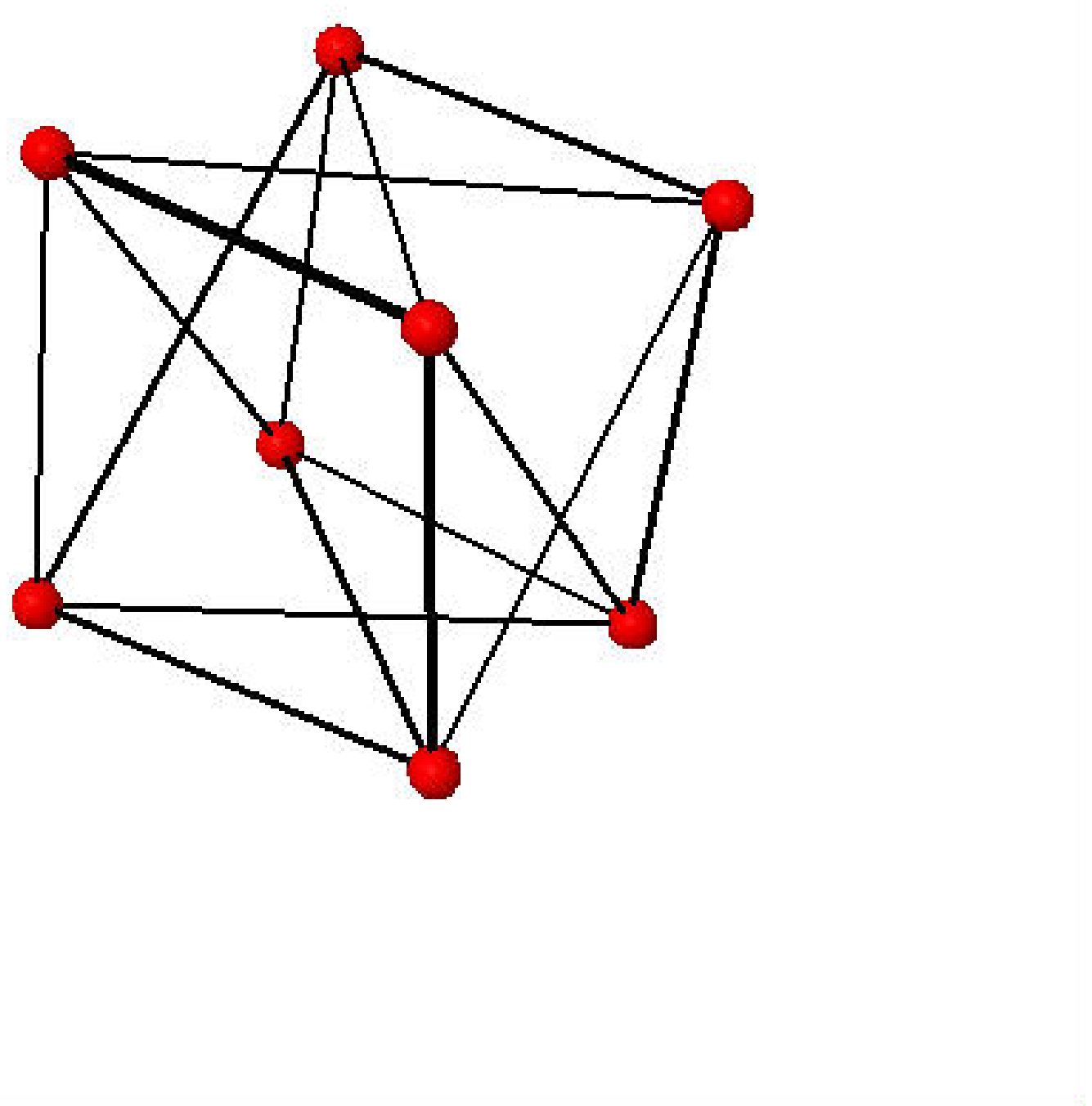}
\caption{Structure produced when all sites are active. This is the most highly connected structure that can be produced from this model.}
\label{connected}
\end{figure}

We show several different structures we have produced by numerical simulation in order to show the variety of topologies our scheme can produce. All show the structure produced and the grid of active sites required to produce it. The number of timesteps was 10 for all of the examples. We can see the creation of a cluster state for universal quantum computing in figs.~\ref{grid1}, \ref{grid2} and \ref{grid3}. In fig.~\ref{grid5} we see additional links from a central structure. These could be used as ancilla qubits for algorithmic or error correction purposes. Finally in figs.~\ref{grid4} and \ref{grid6} we see the initial unit cell of both cubic and and hexagonal lattices. If all sites are active we obtain the most highly connected state possible in this scheme. This is a cube with its corners connected diagonally as shown in fig.~\ref{connected}. This is in essence a superposition of all other structures that could be produced. Other more exotic structures can be created using a different combination of active sites. 

\subsection{Drawbacks}

 \begin{figure}[!tb]
\centering
\includegraphics[scale=0.35]{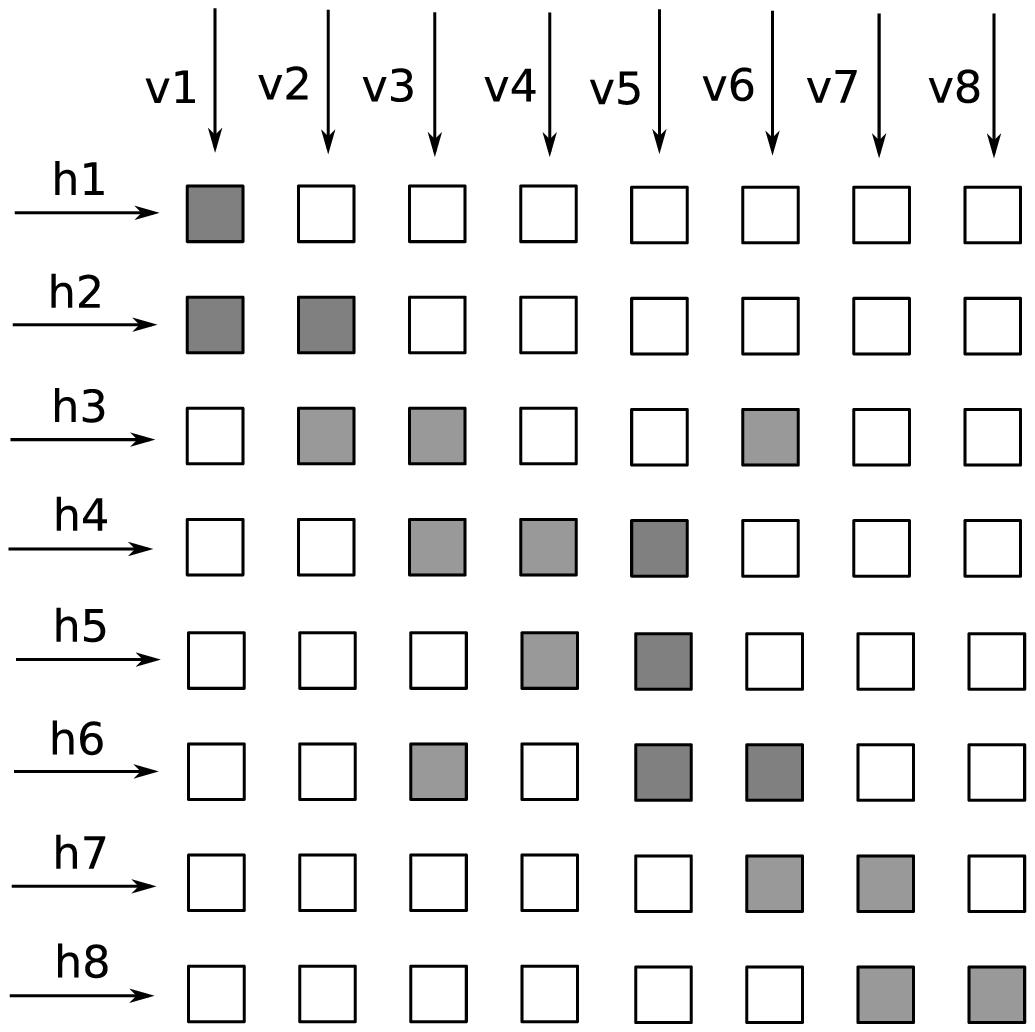}
\includegraphics[scale=0.27]{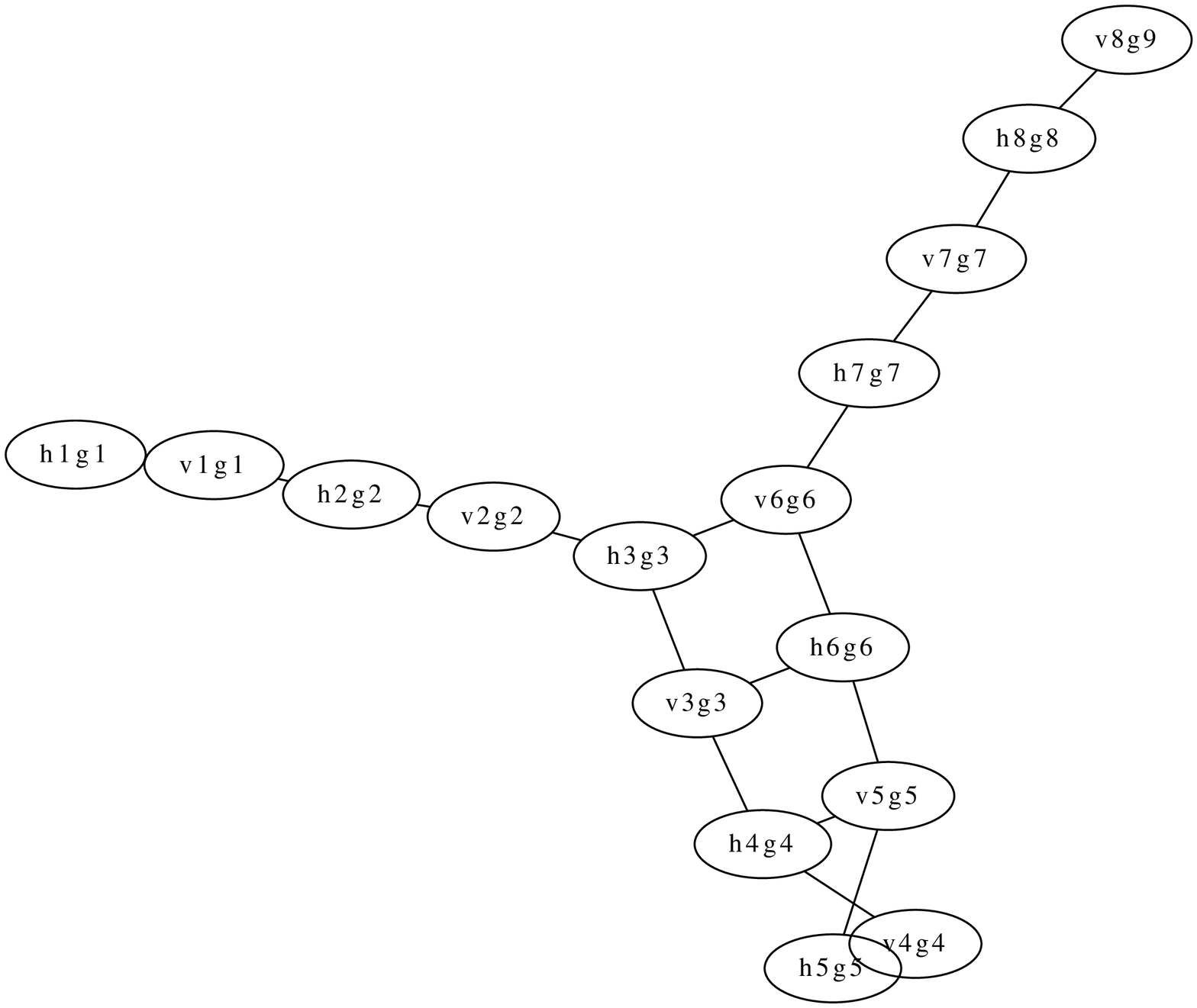}
\caption{Grid of active sites and structure formed for the pattern in fig.~\ref{grid2} repeated and `linked' together. If the number of timesteps was increased then more of the structure would link together at each timestep. The grid has two patterns which would create a 1D lattice, the skew diagonals link the two to form the two depth 2D lattice.}
\label{repeatlink}
\end{figure}

 \begin{figure}[!htb]
\centering
\includegraphics[scale=0.25]{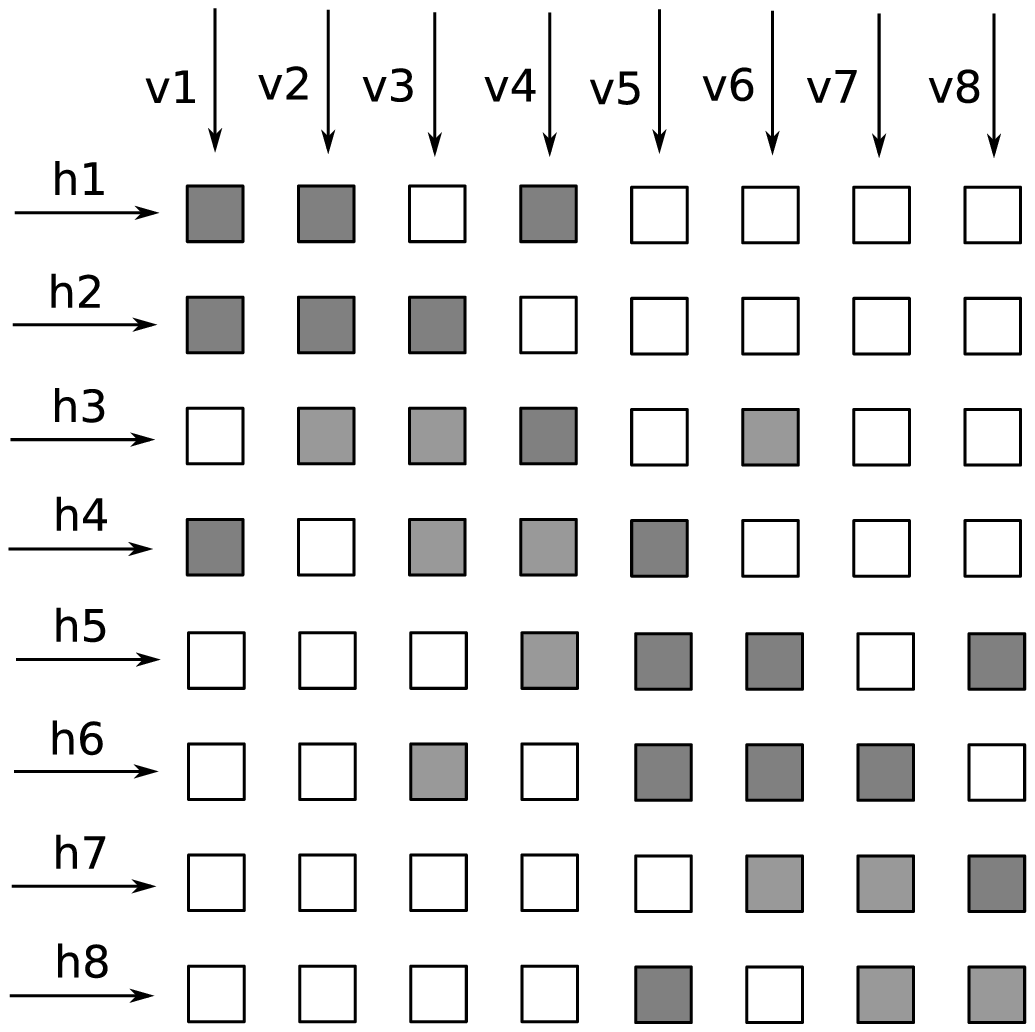}
\includegraphics[scale=0.25]{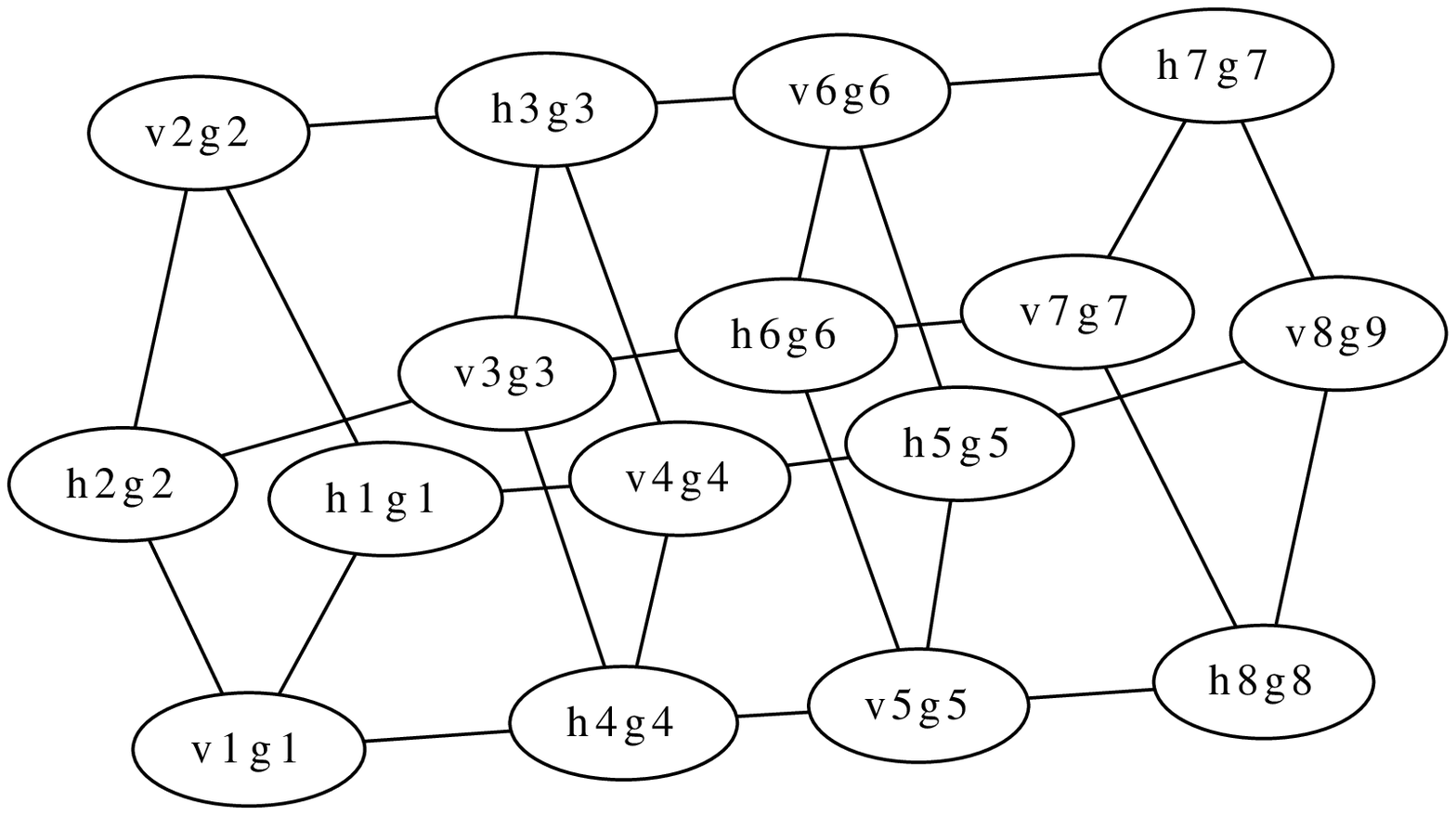}
\caption{Grid of active sites and structure formed for the pattern in fig.~\ref{grid5} repeated and `linked' together. The grid has two patterns which would create cubes, the skew diagonals link the two to form a `chain' of cubes.}
\label{repeatcube}
\end{figure}

In the examples we have shown it is clear that each structure created can only have a maximum of eight vertices due to the size of the grid. Obviously the grid is not static in size, it can be be a square of any dimension $N$, $\sqrt{N} \times \sqrt{N}$. We find that as the grid increases in size, the number of vertices in each structure also increases. The number of vertices present in any structure created can only be a maximum of $2\sqrt{N}$. Similarly, the maximum number of edges a vertex can form is $\sqrt{N}$, depending on how many active sites a vertex passes through. This restricts both the size and the topology of the structures we are able to create. However, as the grid increases in size we do not have to stick to a specific pattern across the entire grid. Another option is to repeat an existing one on the diagonal. For example, if we have an 8 x 8 grid we could have any one of the patterns from figs.~\ref{grid1} to \ref{grid6} repeated twice on the diagonal. This will create additional structures of the same form which could then be linked to form larger structures. This can be achieved by activating the adjacent skew diagonal elements between the repeated pattern. This is shown in figs.~\ref{repeatlink} and \ref{repeatcube} and it is clear that any of the previous structures described could be linked in this way. The only thing that would limit the number of structures we could link together would be the size of the grid that could be constructed. However, as the grid is expanded in this way there are many inactive sites meaning it will probably not scale particulary well in a physically realisable device.

\section{Extended Scheme}
\label{sec:exscheme}

We now modify our scheme to allow the creation of larger structures using the same initial grid. We show some of the useful states we can produce (by numerical simulation).

\subsection{Scheme}

\begin{figure}[!tb]
\begin{minipage}{\columnwidth}
    \centering
	\includegraphics[scale=0.5]{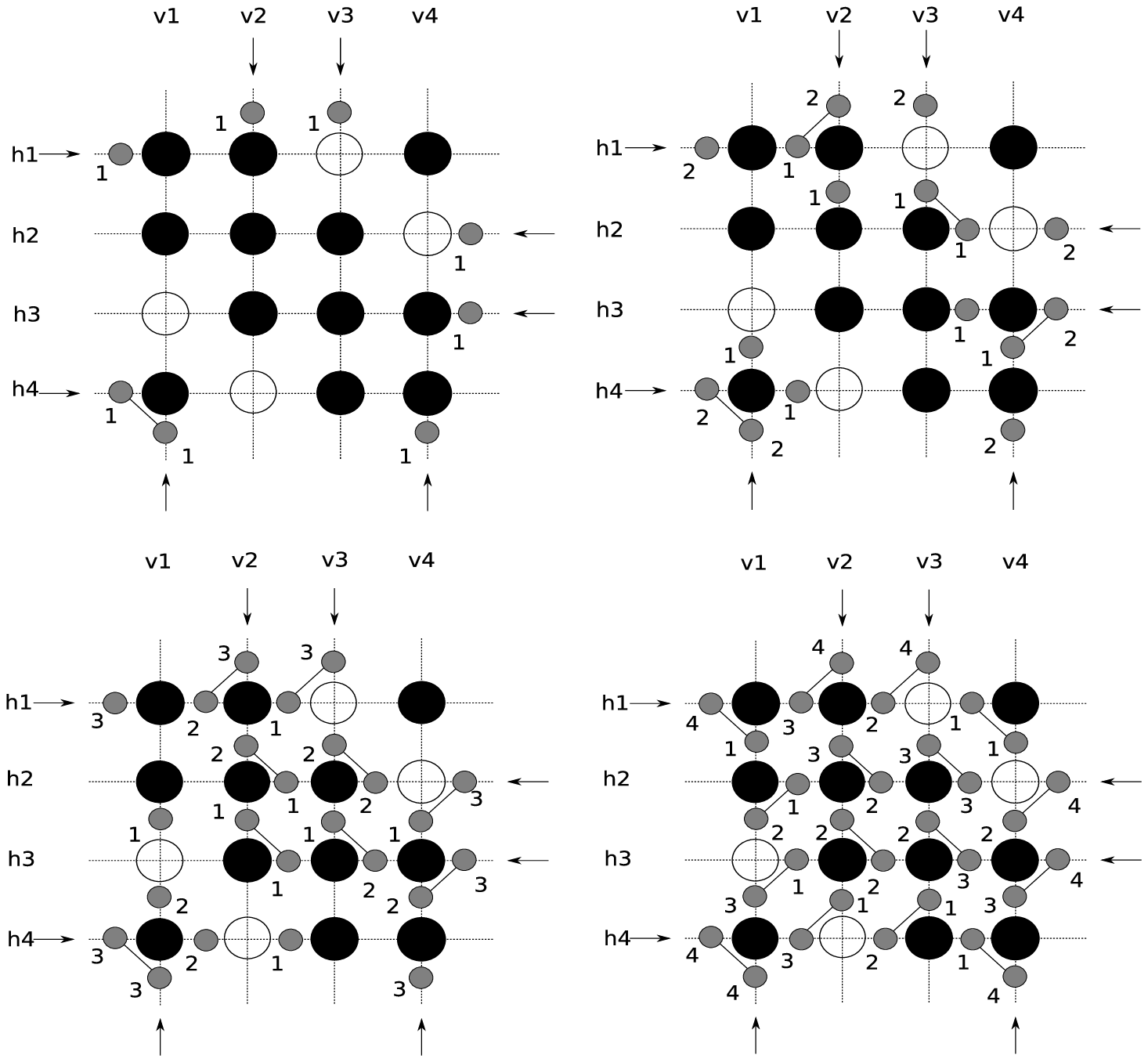}
	\caption{First four timesteps of a 4x4 grid as it 'fills up'. The vertices enter in the direction of the arrows. The black circles indicate active collision sites wheras white indicates the site is switched off.}
    \label{fillgrid}
\end{minipage}
\end{figure}

In order to solve some of the drawbacks in the initial scheme, we amend it slightly allowing the construction of much larger structures. In the extended scheme, we allow the vertices (qubits) to enter from either side of the grid in any pattern. This is shown in fig.~\ref{fillgrid} which shows the grid `filling' up for the first few timesteps. Due to the change in where and when the vertices meet, we find the structures created are in effect `infinite' in length, the only limiting factor is the number of timesteps the system is run for (assuming we can keep the system coherent for this time). This would allow any number of computational steps to be performed on the qubits. If we stick to creating just a 2D lattice (cluster state) then the size of the grid increases linearly with the depth of the cluster produced. This is clearly shown next where we show some of the examples of states produced, again by numerical simulation. The size of the grid compared to the structure produced is now much smaller than the original scheme. 

\begin{figure}[h!bt]
\centering
\includegraphics[scale=0.08]{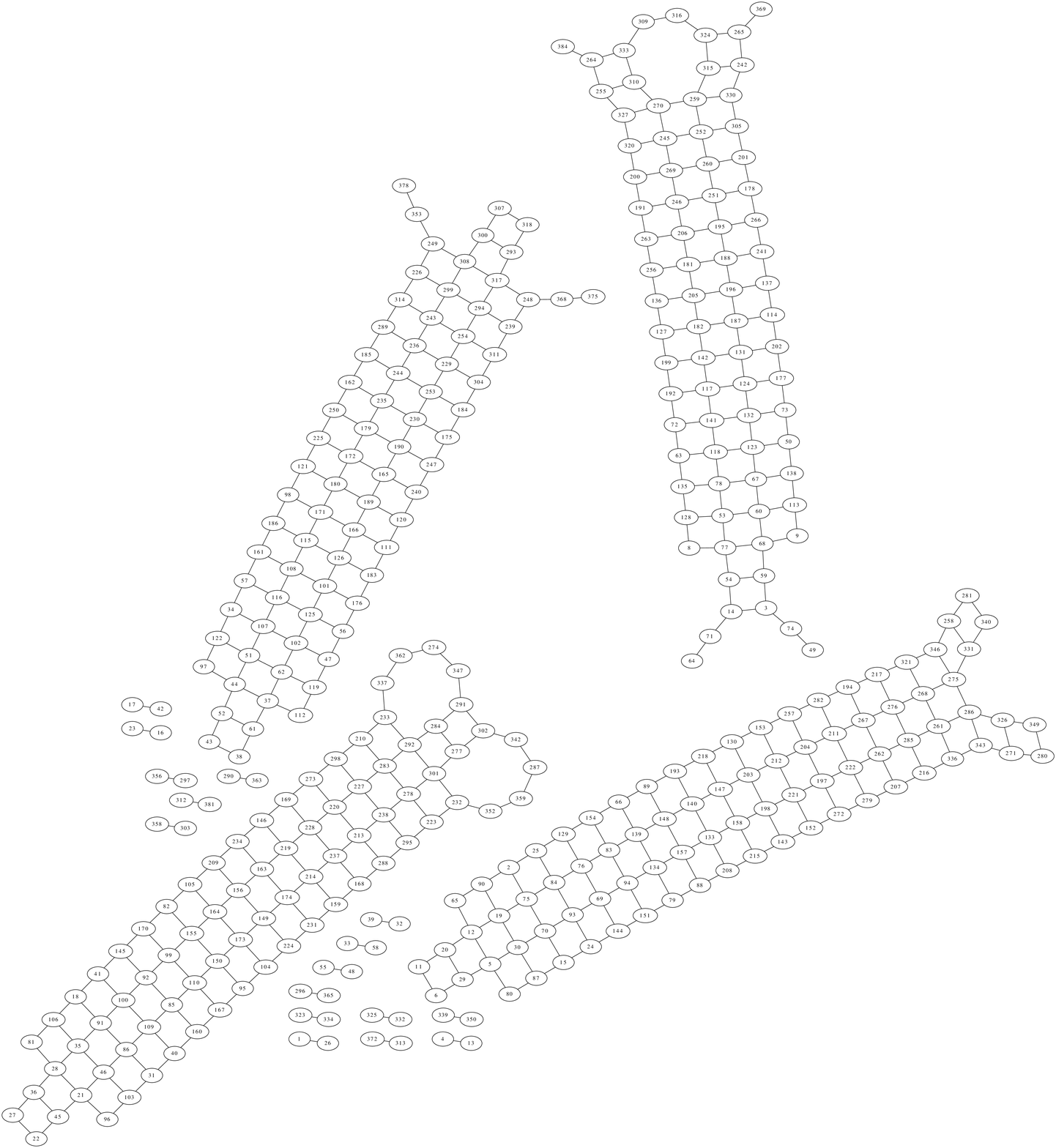}
\caption{An example of the four structures created. The structures show the incomplete parts at the start and end.}
\label{connections}
\end{figure}

Allowing the vertices to enter the grid in this fashion means that only four copies of the same structure are obtained. As is shown by the examples, the structures are much more similar in comparison to the original scheme. As only four structures are produced for any grid pattern it means that the connections are formed much faster. This ensures much of the structure is always complete and it is just the start and end which has less connections. This is shown in fig.~\ref{connections} and is due to two factors. The first is due to the grid filling up at the start of the generation of the cluster which can be solved by starting the computation $\sqrt{N}$ steps later. The second is at the end of the structure which is incomplete as some of the vertices have not passed through the entire grid when we reach the number of required timesteps. This can be solved by ensuring the number of timesteps is an additional $\sqrt{N}$ than required for the computation. Therefore, overall we have a constant overhead of $2\sqrt{N}$ timesteps to add to any computation.

Our initial motivation was to generate a scaleable scheme for cluster state generation. We also found that we could easily create much more interesting structures as in the original scheme. If certain sites are activated we find the structure can loop and join itself again creating 3D structures and `rings'. Other options are to repeat the basic pattern of the grid down the diagonal and then link them together. This could be used to form a `ring of rings' for example. One interesting thing about copying the pattern in this sense is that we can create a `ring' of any number of other structures. These individual structures can also have any number of vertices linked together.

\subsection{States produced}

We show several different structures we have produced by numerical simulation in order to show the variety of topologies our extended scheme can produce. All show the structure produced and the grid of active sites required to produce it.

\begin{figure}[h!bt]
\centering
\includegraphics[scale=0.6]{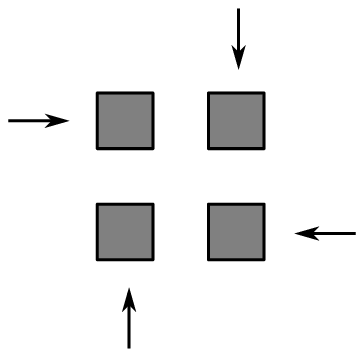}
\includegraphics[scale=0.2]{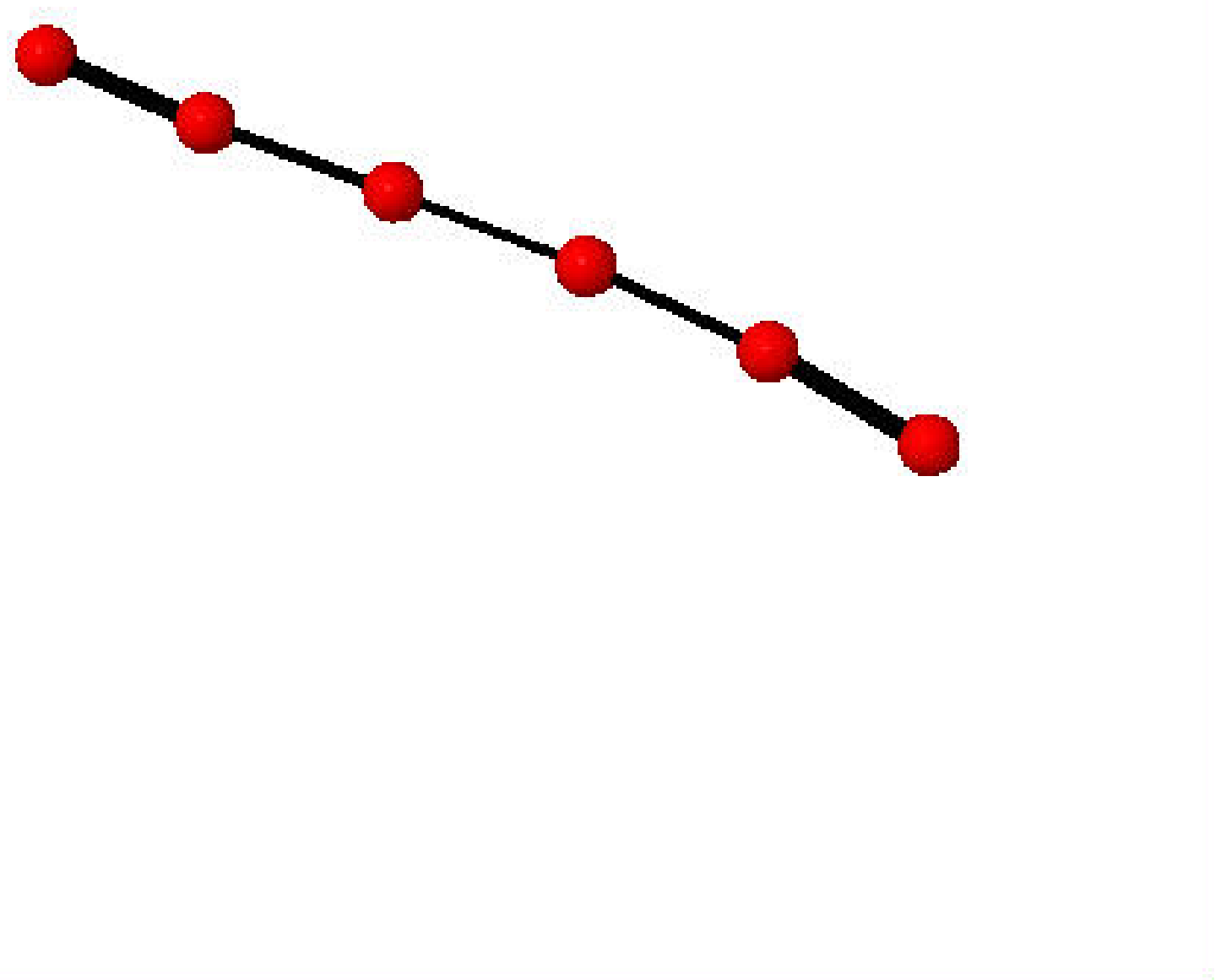}
\caption{Grid of active sites and structure produced. 1D cluster with no restricted length.}
\label{extend1}
\end{figure}

\begin{figure}[h!bt]
\centering
\includegraphics[scale=0.6]{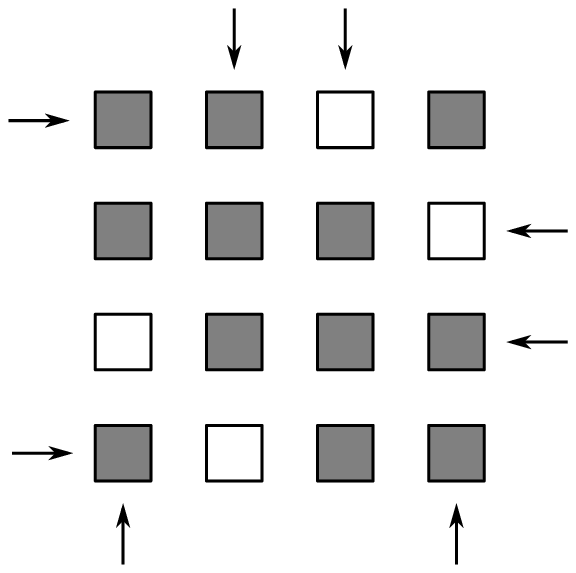}
\includegraphics[scale=0.2]{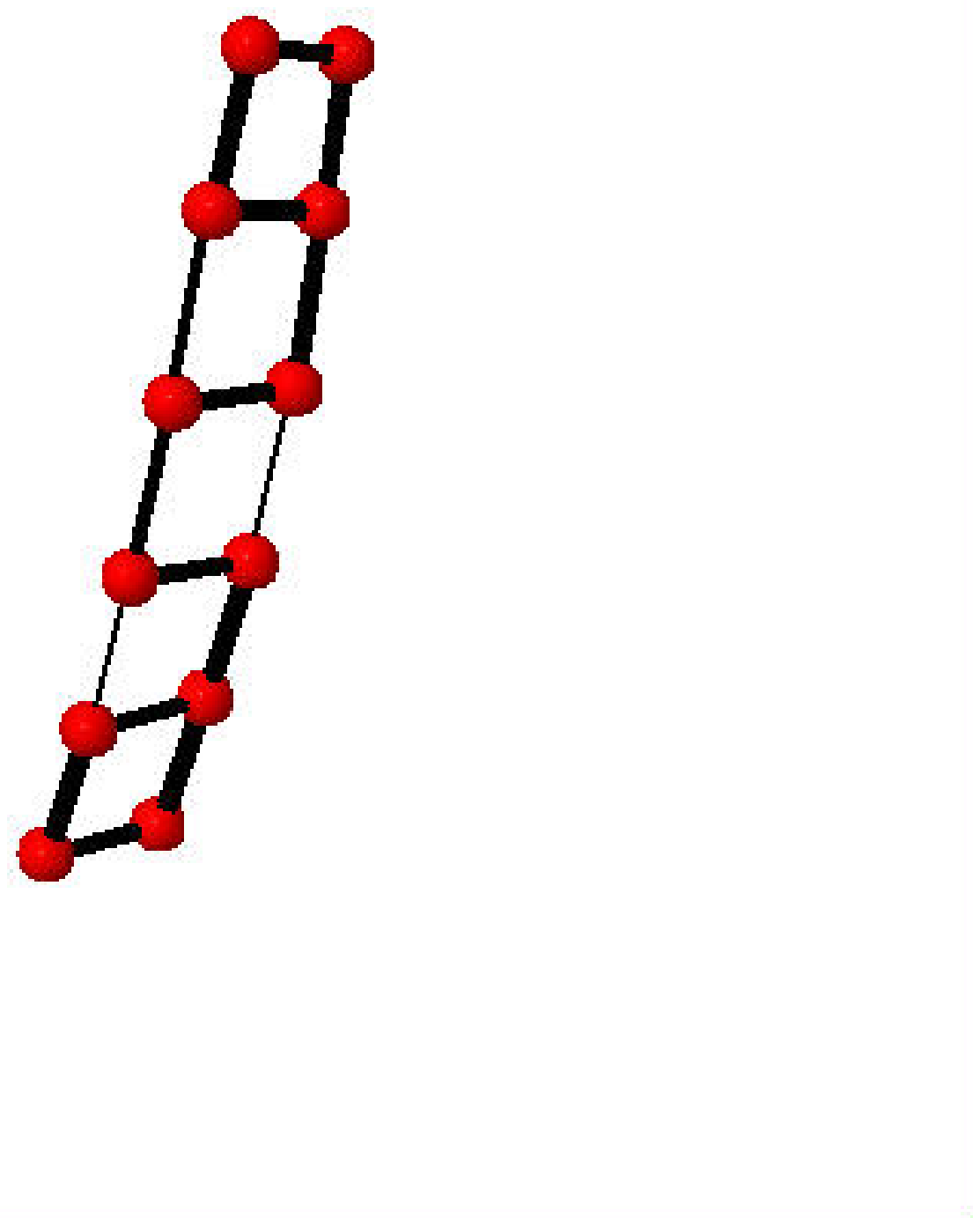}
\caption{Grid of active sites and structure produced. 2D cluster of depth two with no restricted length.}
\label{extend2}
\end{figure}

\begin{figure}[h!bt]
\centering
\includegraphics[scale=0.25]{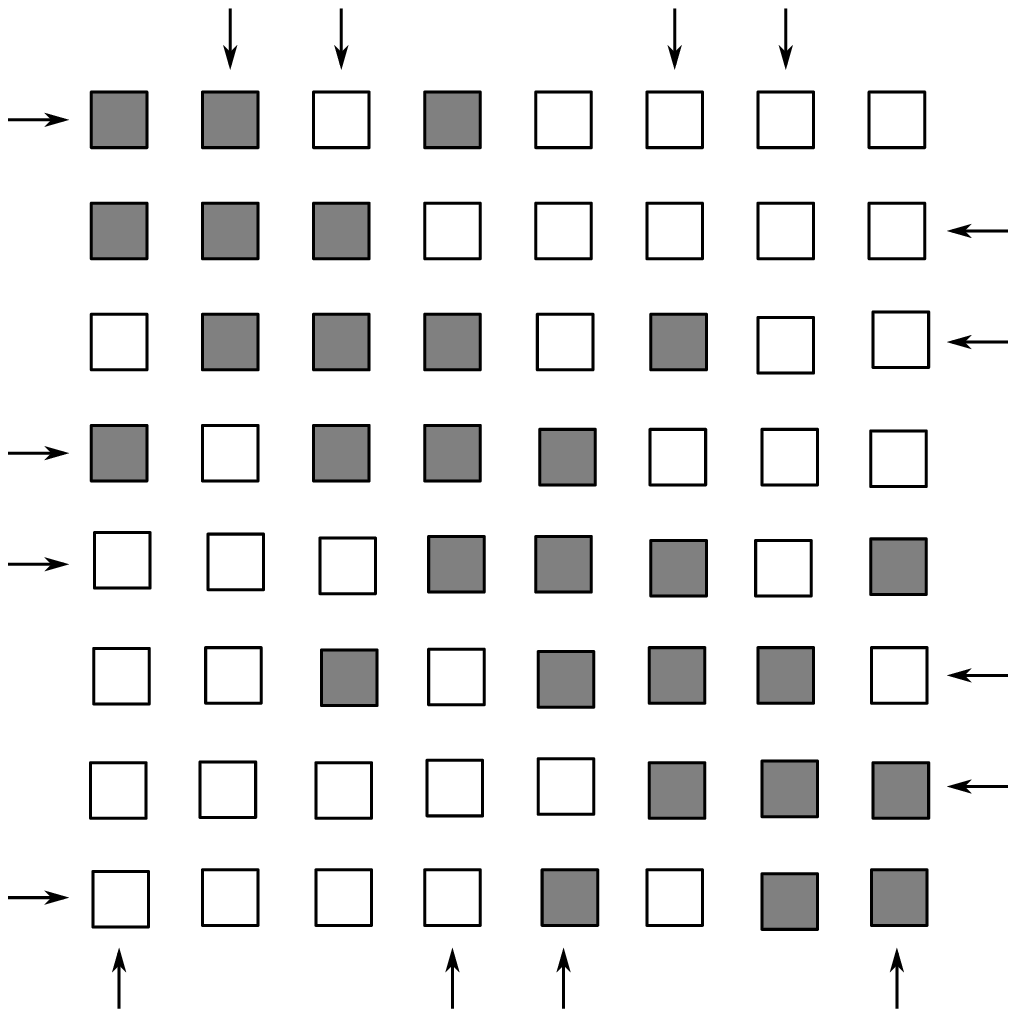}
\includegraphics[scale=0.15]{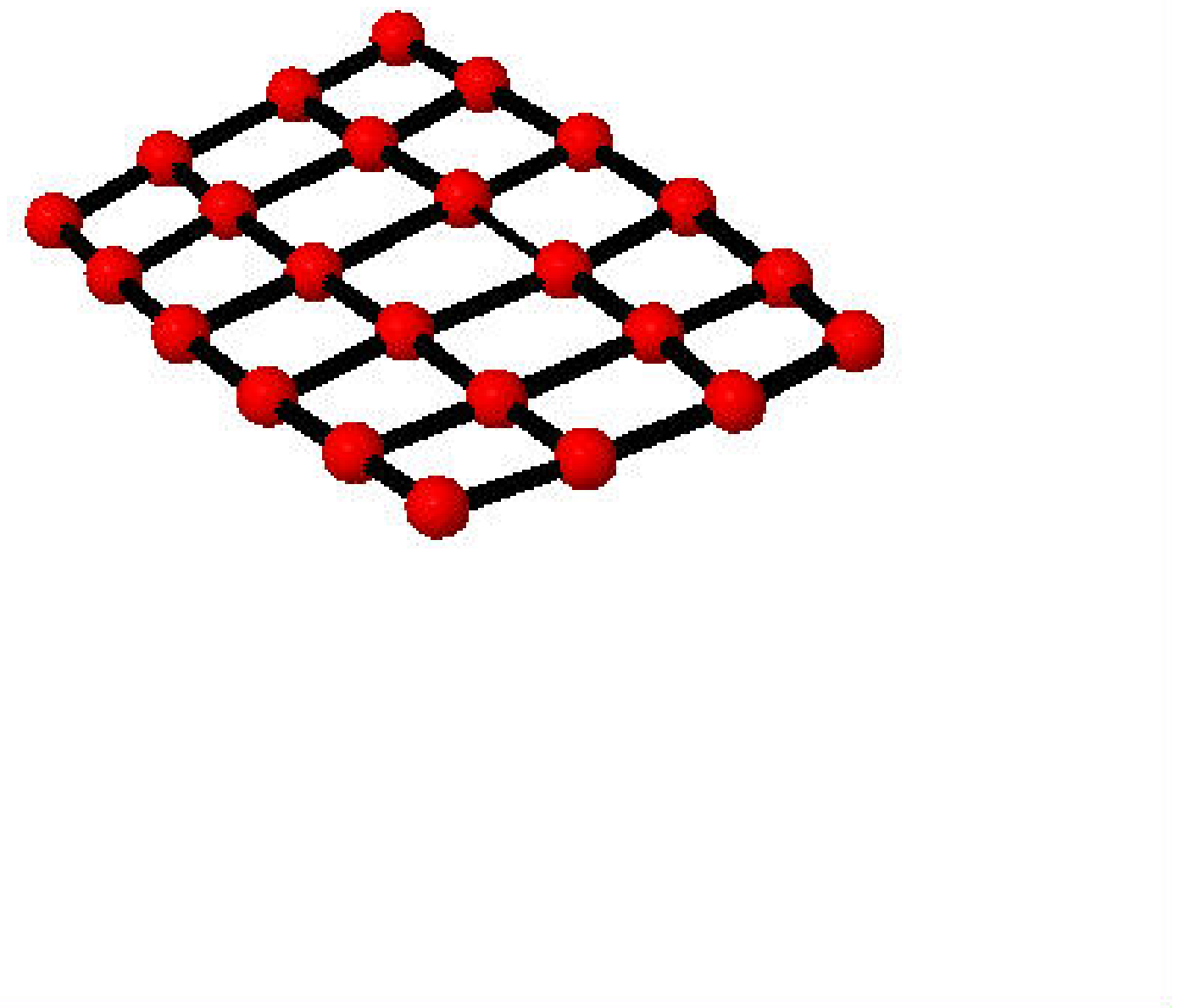}
\caption{Grid of active sites and structure produced. 2D cluster of depth four with no restricted length.}
\label{extend3}
\end{figure}

\begin{figure}[h!bt]
\centering
\includegraphics[scale=0.3]{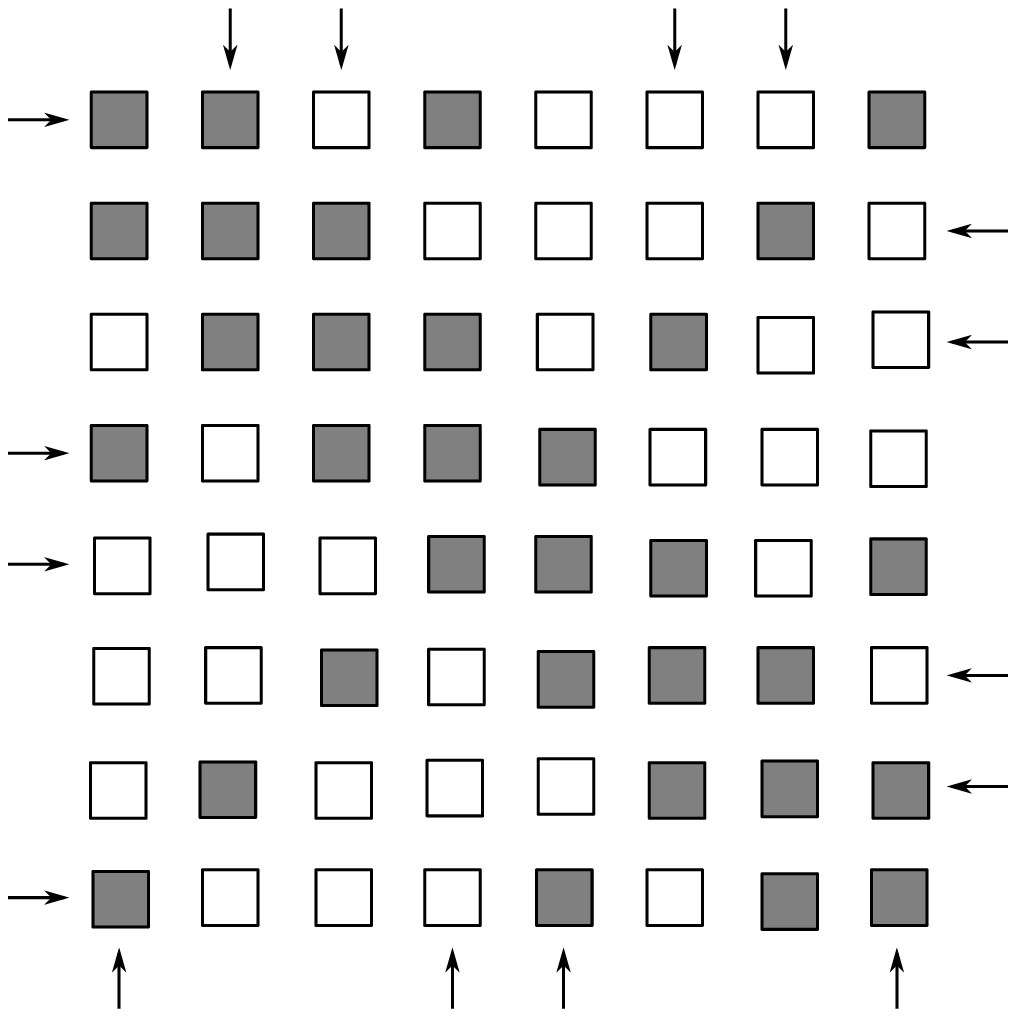}
\includegraphics[scale=0.2]{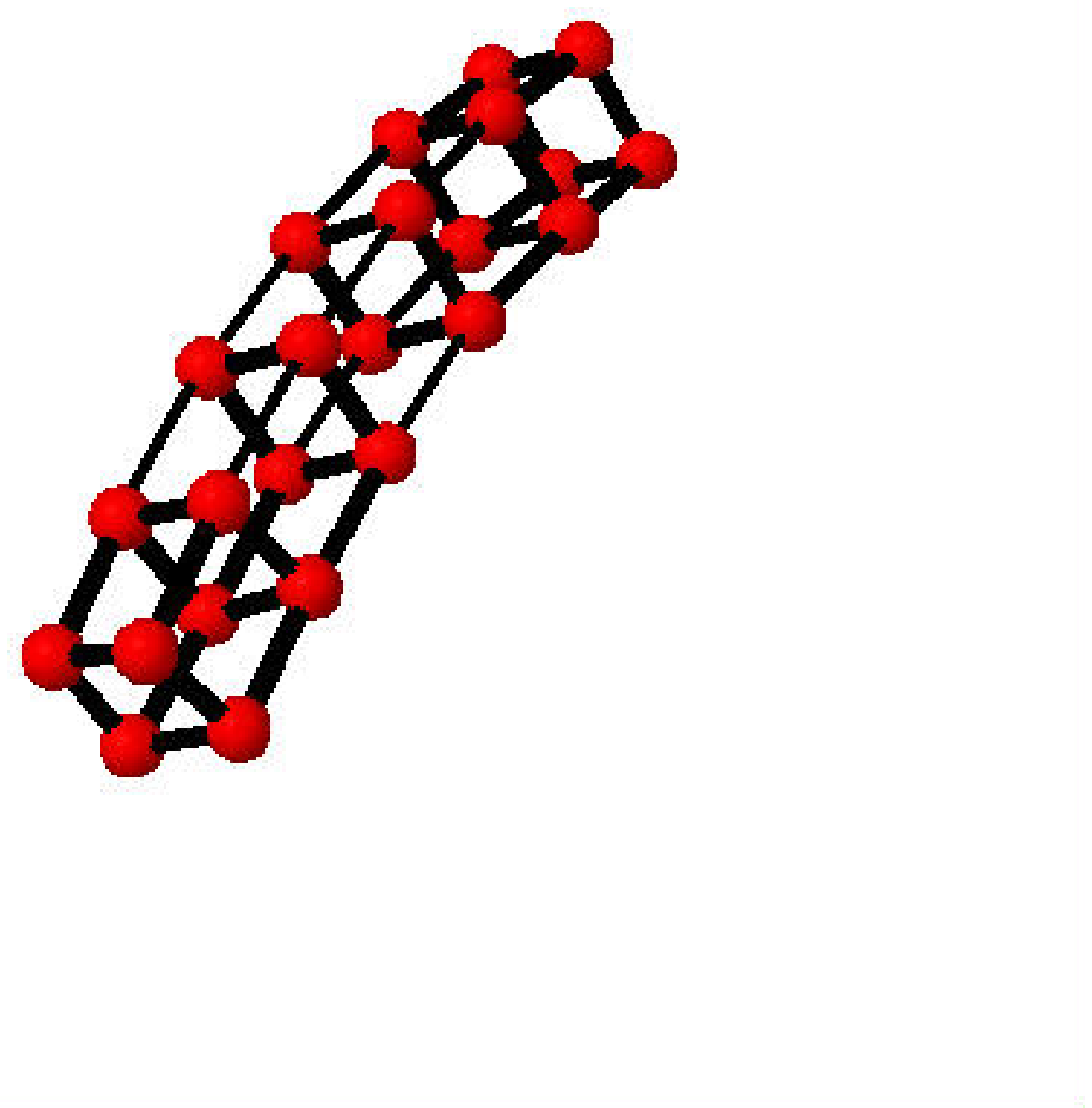}
\caption{Grid of active sites and structure produced. 2D cluster of depth four looped round to form a cube of no restricted length.}
\label{extend4}
\end{figure}

\begin{figure}[h!bt]
\centering
\includegraphics[scale=0.18]{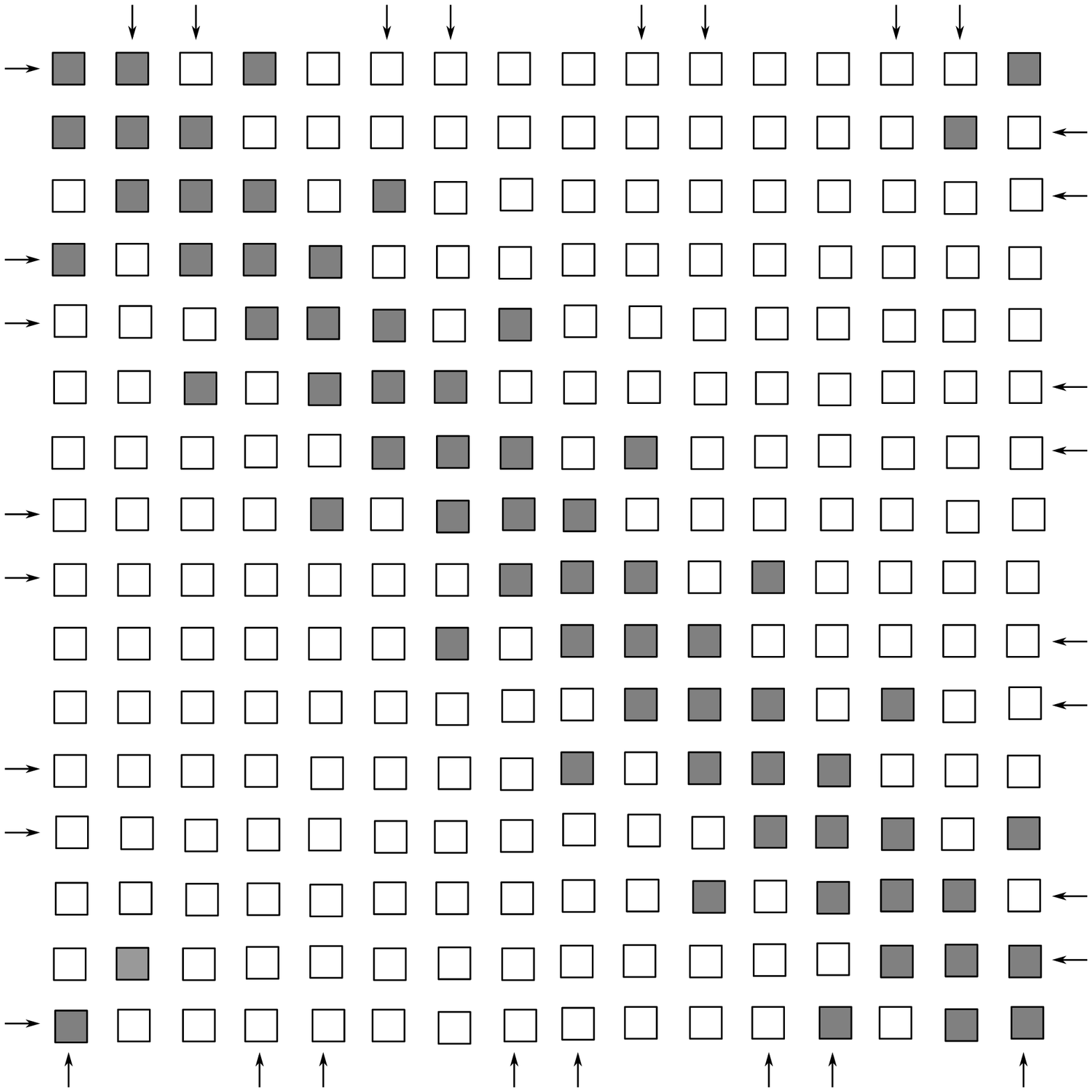}
\includegraphics[scale=0.28]{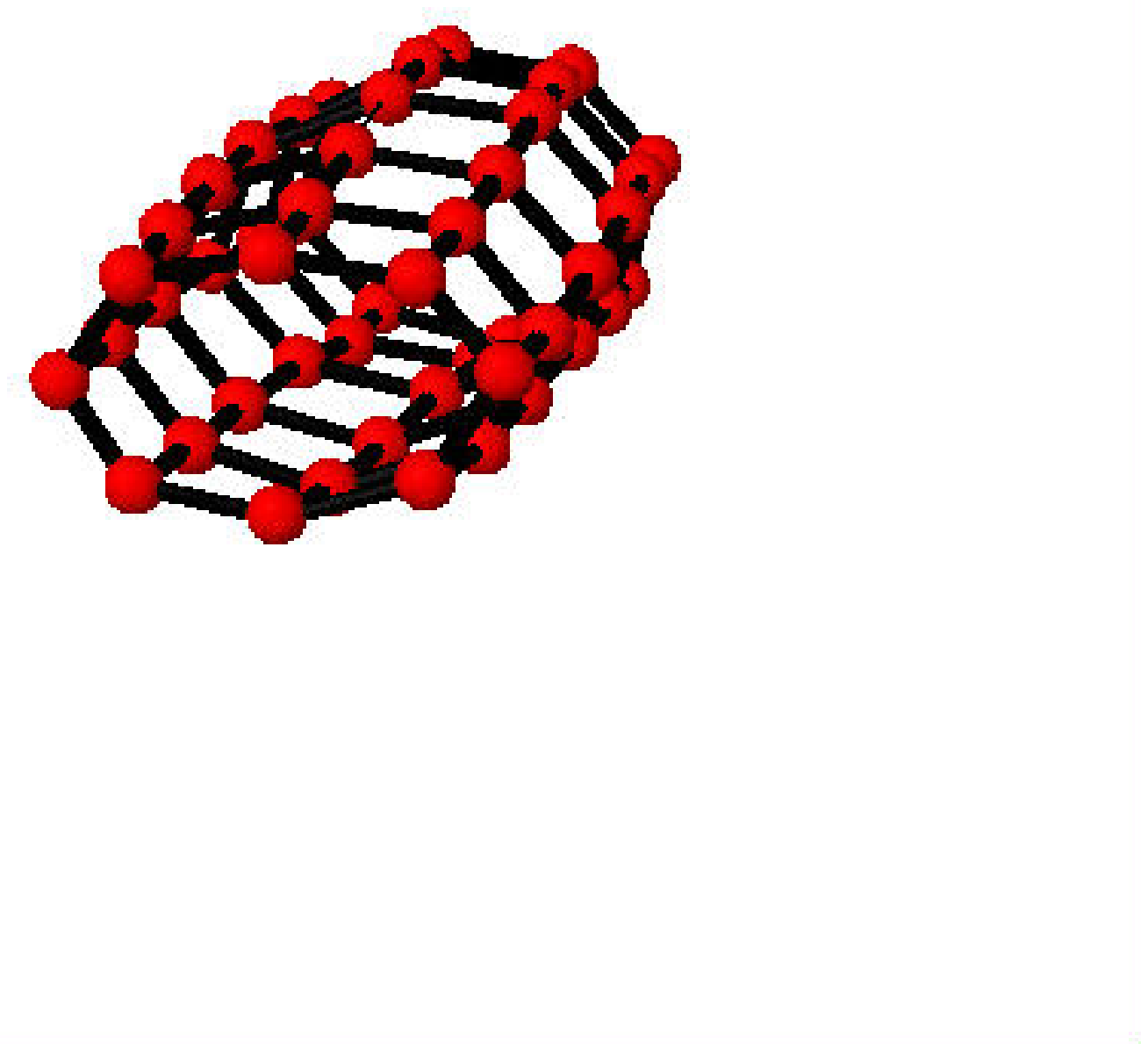}
\caption{Grid of active sites and structure produced. 2D cluster of depth 8 looped round to form an octagon of no restricted length.}
\label{extend5}
\end{figure}

We can see the creation of a cluster state for universal quantum computing in figs.~\ref{extend1}, \ref{extend2} and \ref{extend3} and can clearly see the linear scaling of the grid with cluster depth. Figures \ref{extend4} and \ref{extend5} show how activating specific sites allows depth four or eight clusters to loop around to form a cube or octagon respectively. The 3D structures shown in figs.~\ref{extend4} and \ref{extend5} can be repeated in a larger grid and then linked together in a similar way. This then achieves a `ring of rings' topology where the number of vertices in either ring is just dependent on the size of the initial grid.

\section{Discussion / Further Work}
\label{sec:conc}

We have presented a scheme to allow the creation of graph states. Our scheme could be applied to various architectures, however we do note its relevance to the cavity QED scheme previously mentioned, \cite{varcoe06a}. In this way we envisage the vertices as atoms and the edges of the graph as entanglement between these atoms. This architecture would seem to lend itself to quantum information processing applications. The basic cluster state produced by the extended scheme is a universal resource for measurement based quantum computing. The scaling here is better than many other schemes that have been proposed. We only need to double the size of the grid to get a structure of double the depth (double the qubits in the cluster state). It is clear that doing this will create many inactive collision sites and as such is wasteful. However, if experimentally implemented these inactive sites need not be implemented, instead just creating the active zones and controlling the timing of the atoms is sufficient. Obviously, maintaining the timing and coherence of the qubits during the computation would still represent a significant challenge. As the computation can be run for an arbitrary time, the length of the structure is in effect `infinite' as long as the system remains coherent. The unit cell for topological error correction can also be created (with some Z measurements to remove qubits).

In the scheme, we have assumed that when two vertices pass each other at an active site a full link or edge is always formed. This is an ideal case and we intend to amend our numerical simulations to include a probability of an edge forming. We would imagine there to be some critical probability similar to percolation theory where the structures are formed / not formed. This could mean the structures could be used in percolation problems when studying stuctures with broken or missing links. We notice that as there are multiple copies of the same structure it is unlikely that the same bonds will form in each one if there is chance of an error. As such we may be able to use some form of `majority rules' or entanglement distillation scheme, \cite{cirac07a}, to ensure we have a complete structure. This would however remove many of the additional structures that are in effect created for `free'.

The additional structures could easily be used to our advantage. This could be one of two options - multiple copies running either the same computation or different parts of a program. If we ran the same program on all the copies this would give a higher probability of success if there was a possibility of error as discussed above. It would also mean a probabilistic algorithm would have to be run less times as we would be in effect running it four times in one run. 
The other option would be to use each structure as a `thread' in a multithreaded quantum computer. Activating specific sites in the grid at certain times would then allow the connection of the structures temporarily to allow communication between threads. This would be much more complex than the previous option but could also give benefits such as speed for example. The main problem with the multithreading idea would be communication between threads. This would be in the timing of the links between structures to pass information from one structure to another. It would also require longer coherence times as the states may need to be `stored' temporarily if communication is blocked with another structure to avoid a read-write error between threads in the same way as classical multithreading.

We intend to extend this work to provide a more physical setting in which we can discuss errors and implementation more thoroughly. The scheme we present here can not produce every structure as the vertices (qubits) cannot be directed from output to input as in \cite{devitt07a}. We will address this in a specific architecture. Another possibility of the scheme is useful applications of the more exotic structures shown in the extended scheme. This would most likely be in the area of topological error correction mentioned briefly above. The unit cell can be created which is the basis of the error correcting schemes introduced by Raussendorf and Harrington \cite{raussendorf07a}. An extension of error correction in this scheme would be to use multiple qubits in order to encode one logical qubit. This redundancy allows for fault tolerance but how many qubits we could use would obviously depend on how many qubits could physically be realisable.

\paragraph{Acknowledgments:}
NL is funded by the UK Engineering and Physical Sciences Research Council.
BV is supported by the UK Engineering and Physical Sciences Research Council through an Advanced Fellowship GR/T02331/01 and Grant GR/S21892/01

\bibliography{nwc}{}

\end{document}